\documentclass[11pt]{article}

\usepackage{geometry,setspace}

\usepackage{natbib}

\usepackage{makeidx, siunitx, ragged2e}
\usepackage{booktabs,color}
\usepackage{amsfonts,bm}
\usepackage{amsmath}
\usepackage{amssymb}
\usepackage{graphicx}
\usepackage{rotating}
\usepackage{multirow}%
\usepackage{multicol}
\usepackage{subfigure}
\usepackage{rotating}
\usepackage{tikz}
\usepackage{pdfpages}
\usepackage{ulem}
\usepackage{booktabs}
\usepackage{lscape}
\usepackage{caption}
\usepackage{lipsum}
\usepackage{longtable}
\usepackage{booktabs}
\newcommand{\E}{\mathbb{E}}

\newcommand{\nc}{\normalcolor}

\begin{document}

\title{The Financial Market of Indices of Socioeconomic Wellbeing}

\author{
	Thilini V. Mahanama\thanks{Department of Industrial Management, University of Kelaniya, Sri Lanka, thilinim@kln.ac.lk (Corresponding
		Author).}
	\and 
	Abootaleb Shirvani\thanks{Department of Mathematical Science, Kean University, Union, NJ, U.S.A, ashirvan@kean.edu.}
	\and 
 	Svetlozar Rachev\thanks{Texas Tech University, Department of Mathematics
	\& Statistics, Lubbock TX 79409-1042, U.S.A., Zari.Rachev@ttu.edu}
}
\date{}
\maketitle


\begin{abstract}


The financial industry should be involved in mitigating the risk of downturns in the financial wellbeing indices around the world by implementing well-developed financial tools such as insurance instruments on the underlying wellbeing indices.
We define a new quantitative measure of the wellbeing of a country's population for those countries using 
the world development indicators provided by the World Bank.
We monetize the indices of socioeconomic wellbeing, which serve as “risky assets,” 
and consequently develop a global financial market for them, which serves as a “market of indices of socioeconomic wellbeing.” 
Then, we compare the wellbeing of different countries using financial econometric analysis and dynamic asset pricing theory.
We  provide the optimal portfolio weight composition along with the efficient frontiers of the wellbeing socioeconomic indices with different risk--return measures.
We derive insurance instruments, such as put options, 
which allow the financial industry to monitor, manage, and trade these indices, creating the funds for insurance against adverse movements of those indices. 
%
Our findings should help financial institutions to incorporate socioeconomic issues as an additional dimension to their ``two-dimensional" risk--return adjusted optimal financial portfolios.

\end{abstract}



\section{Introduction}\label{sec:Introduction}



``The greatest evils and the worst of crimes is poverty; our first duty, a duty to which every other consideration should be sacrificed, is not to be poor.” --- Bernard Shaw\\

The experimental approach proposed by Esther Duflow, Abhijit Banerjee, and Michael Kremer to alleviating global poverty by improving education or health in the developing world or addressing some of the other problems affecting the very poor was awarded the Nobel Memorial Prize in Economic Sciences in 2019.
In \citep{reinsdorf2020measuring}, the International Monetary Fund (IMF) discusses the distributional indicators of income, consumption, and wealth in a national accounts framework for categorizing indicators of wellbeing and economic welfare.



\cite{ahmad2021quantifying} discusses the importance of proposing an index that captures the level of wellbeing of the average inhabitant of a country, as this will allow policymakers to benchmark their performance. 
For instance, the United Nations development programme published the human development index, using 
long and healthy life, being knowledgeable, and have a decent standard of living as indicators \citep{sagar1998human}.
The organisation for economic co-operation and development (OECD) is an initiative pioneering the development of economic indicators which better capture multiple dimensions of economic and social progress \citep{OECD}.
The OECD better life index reflects the current socioeconomic status of a country based on its measures related to housing, income, jobs, education, the environment, quality of social support network, governance, health, life satisfaction, safety, and work--life balance.


We want to have a global view of the wellbeing of a society, applying the tools of dynamic asset pricing theory.
We believe that the financial industries should be involved in evaluating and mitigating the risk of adverse movements of the indices
using the methods developed for financial portfolio insurance.
Similar work has been  done in environmental, social, and governance (ESG) and socially responsible investing (SRI) \citep{matos2020esg}. 
In this paper, we consider a more general problem by 
adding to the risk--return profiling of financial institutions an additional socioeconomic dimension,  namely \nc the wellbeing of the society in general.
This approach should incorporate the ESG factors as a subset of all socioeconomic factors contributing to the wellbeing of the society.

\cite{trindade2020socioeconomic} introduced an index  of socioeconomic wellbeing by constructing a historical national wellbeing
index based on the mood of the population of the US.
They issued marketable financial contracts, such as options and futures, based on financial econometric modeling and dynamic asset pricing theory.
These findings will provide early warnings for policy makers and private agents about potential future market downturns in the mood of the population.


The wellbeing of a country not only depends on the mood of the population or political misconduct, but it also should involve the business community. 
A poor socioeconomic wellbeing index should be viewed as a poor business.
The goal of this paper is to create a global financial market model of wellbeing indices, and thus allowing the financial industry to monitor, analyze, forecast, and actively be involved in managing the systematic and idiosyncratic risks of those socioeconomic indices and  furthermore to \nc provide financial instruments to mitigate potential future adverse movements of the indices.
In this paper, we define a new quantitative measure of wellbeing within the framework of modern asset pricing theory.
The goal of our approach is to monetize the wellbeing of socioeconomic indices in order to arm financial institutions with the necessary financial instruments to assess and manage the risk of adverse movements of those indices.

%

We construct wellbeing indices for countries categorized by gross domestic product per capita, GDP, namely, the United States, Australia, Brazil, China, Germany, India, Japan, South Africa, and the United Kingdom.
We incorporate eight world development socioeconomic indicators:  the Gini index, unemployment, life expectancy at birth, gross national income, consumer price index, population, foreign direct investment, and GDP
published by the World Bank  \citep{Data}. 
For each country, we introduce a dollar denominated index
(determining the financial value of the socioeconomic index).
We incorporate the GDP of a country into the socioecoomic index providing the asset price of the wellbeing index as a financial asset.
This allows us to perform a financial econometric forecast of the future behavior of wellbeing indices as financial assets.
Next, we study the financial market of wellbeing indices using portfolio theory with different risk--return measures as required in the Basel II Accord \citep{de2006basel}.
We compare the wellbeing of the socioeconomic indices of different countries using an econometric analysis and classical asset pricing approaches, such as portfolio efficient frontier with different risk measures, again as required by the Basel II Accord.
We derive financial put options which can be viewed as insurance instruments on the socioeconomic indices for the financial industry in assessing and managing those adverse risks.

%

From the existing literature, it is clear that all existing wellbeing indices are static, that is, there is no underlying time series econometric model to be used by the financial industry to forecast, price, and ultimately trade those indices as part of their financial portfolios.
We are not aware of any academic research  involving financial instruments designed to assess the downturns of a society.
Therefore, we create the financial market of socioeconomic indices
in order to allow the financial industry to be engaged in the wellbeing of society by assessing and managing the potential future downturns in the wellbeing of a society.
While the ESG financial markets are important components in the financial markets dealing with socioeconomic issues, our goal is more global, encompassing the wellbeing of the society as a whole.
 



As mentioned above, the goal of our paper is to provide a financially sound model for the business community to be involved in managing the risk from the downturns of wellbeing indices.
That can be achieved by trading specialized derivative instruments similar to those used in portfolio insurance, see \citep{agic2017portfolio}.  
For this purpose, we propose new financial instruments to assess the wellbeing of the population of each country and manage the potential adverse movements of that wellbeing.
The financial industry should be able to offer insurance instruments to mitigate and hedge the risk of downturns in the wellbeing indices around the world.



The remainder of this paper's contents are as follows.
First, we construct dollar denominated wellbeing indices for nine countries using 
the eight world development indicators introduced in Section \ref{sec:DataDescription}.
Second, we propose a global US dollar denominated index  of socioeconomic wellbeing and study the dynamics of the asset prices.
Third, in Sections \ref{sec:TailRisk} and \ref{sec:RR}, we evaluate the impacts of adverse events in each country's index on the global wellbeing index.  
Next, we start the study of the financial market of these indices by employing the tools of dynamic asset pricing theory. 
With this goal, we provide the optimal portfolio weight composition along with the efficient frontier in Section \ref{sec:EF}.
In Section \ref{sec:OP}, we provide an option pricing model for the global socioeconomic wellbeing index.
Finally, we make concluding remarks in Section \ref{sec:DC_CI}.

\section{World development socioeconomic indicators} \label{sec:DataDescription}

The World Bank \citep{WorldBank} compiles world development socioeconomic indicators \citep{Data} that are high quality, and internationally comparable statistics about global development and the fight against poverty. 
This section describes the world development indicators we use for constructing our index.

Our US dollar-denominated wellbeing socioeconomic index mainly incorporates the indicators contributed to income, health, labor, education, economy, and global links.
We use the Gini index to capture the distribution of income (poverty and inequality).
The total population, life expectancy at birth, and unemployment indicators provide dynamics related to the population in terms of health, labor, and education.
Gross national income (GNI), the consumer price index (CPI), and GDP are used as measurements of income and savings, prices, and terms of trade, and the growth and economic structure of a country.
We use foreign direct investment to include financial flows on global links.
The World Bank definitions for the aforementioned world development indicators are given below:

\begin{itemize}
	
\item {Gini index (World Bank  estimate)}\\
The Gini index measures the extent to which the distribution of income (or, in some cases, consumption expenditure) among individuals or households within an economy deviates from a perfectly equal distribution. A Lorenz curve plots the cumulative percentages of total income received against the cumulative number of recipients, starting with the poorest individual or household. The Gini index measures the area between the Lorenz curve and a hypothetical line of absolute equality, expressed as a percentage of the maximum area under the line. Thus a Gini index of 0 represents perfect equality, while an index of 100 implies perfect inequality.
100 – Gini index

\item {Unemployment, total (\% of total labor force) (modeled ILO estimate)}\\
Unemployment refers to the share of the labor force that is without work but available for and seeking employment.

\item {Life expectancy at birth, total (years)}\\
Life expectancy at birth indicates the number of years a newborn infant would live if prevailing patterns of mortality at the time of its birth were to stay the same throughout its life.

\item {GNI per capita, Atlas method (current US\$)}\\
GNI per capita (formerly GNP per capita) is the gross national income, converted to U.S. dollars using the World Bank Atlas method, divided by the midyear population. GNI is the sum of value added by all resident producers plus any product taxes (less subsidies) not included in the valuation of output plus net receipts of primary income (compensation of employees and property income) from abroad. GNI, calculated in national currency, is usually converted to U.S. dollars at official exchange rates for comparisons across economies, although an alternative rate is used when the official exchange rate is judged to diverge by an exceptionally large margin from the rate actually applied in international transactions. To smooth fluctuations in prices and exchange rates, a special Atlas method of conversion is used by the World Bank. This applies a conversion factor that averages the exchange rate for a given year and the two preceding years, adjusted for differences in rates of inflation between the country, and through 2000, the G-5 countries (France, Germany, Japan, the United Kingdom, and the United States). From 2001, these countries include the Euro area, Japan, the United Kingdom, and the United States.

\item {Consumer price index (2010 = 100)}\\
The consumer price index reflects changes in the cost to the average consumer of acquiring a basket of goods and services that may be fixed or changed at specified intervals, such as yearly. The Laspeyres formula is generally used. Data are period averages.

\item {Population, total}\\
The total population is based on the de facto definition of population, which counts all residents regardless of legal status or citizenship. The values shown are midyear estimates.

\item {Foreign direct investment, net (BoP, current US\$)}\\
Foreign direct investment are the net inflows of investment to acquire a lasting management interest (10 percent or more of voting stock) in an enterprise operating in an economy other than that of the investor. It is the sum of equity capital, reinvestment of earnings, other long-term capital, and short-term capital as shown in the balance of payments. This series shows total net FDI. In BPM6, financial account balances are calculated as the change in assets minus the change in liabilities. Net FDI outflows are assets and net FDI inflows are liabilities. Data are in current U.S. dollars.

\item {GDP (current US\$)}\\
GDP at purchaser's prices is the sum of gross value added by all resident producers in the economy plus any product taxes and minus any subsidies not included in the value of the products. It is calculated without making deductions for depreciation of fabricated assets or for depletion and degradation of natural resources. Data are in current U.S. dollars. Dollar figures for GDP are converted from domestic currencies using single year official exchange rates. For a few countries where the official exchange rate does not reflect the rate effectively applied to actual foreign exchange transactions, an alternative conversion factor is used.
\end{itemize}


Whenever necessary, we compute missing data using multiple imputations with principal component analysis \citep{josse2011multiple}. 
We modify the world development indicators to construct the global dollar wellbeing index in Section \ref{sec:DWI}.

\subsection{Constructing a global dollar wellbeing index}\label{sec:DWI}

In this section, we construct a global wellbeing index using 
the eight world development indicators introduced in Section \ref{sec:DataDescription} for 
nine countries: the United States, Australia, Brazil, China, Germany, India, Japan, South Africa, and the United Kingdom.
Based on data availability in the database \citep{Data}, we use the reported data between 1990 and 2020.

\begin{figure}[h!] 	
	\centering
	\subfigure[]{\includegraphics[width=0.45\textwidth]{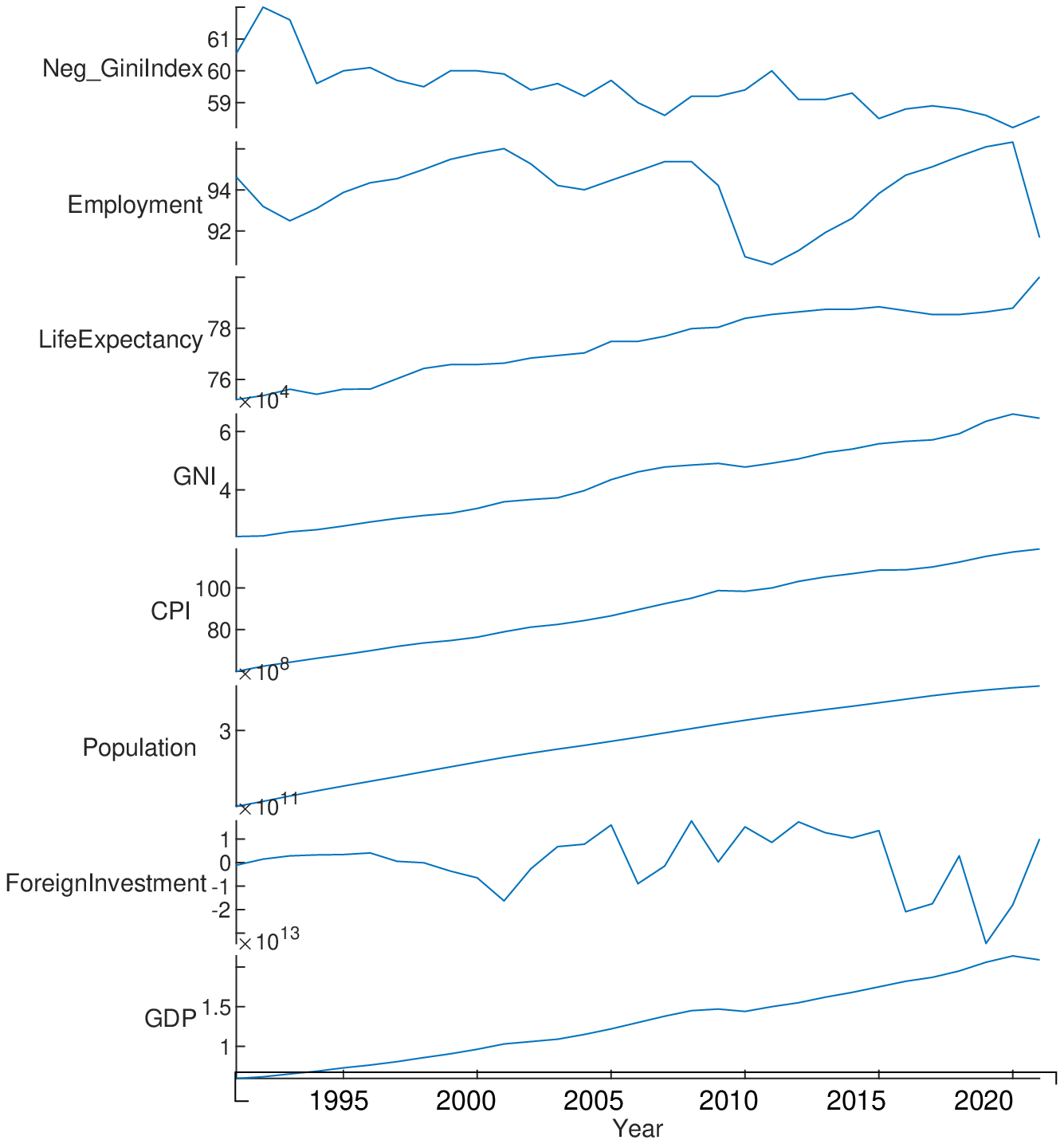}}
	\subfigure[]{\includegraphics[width=0.42\textwidth]{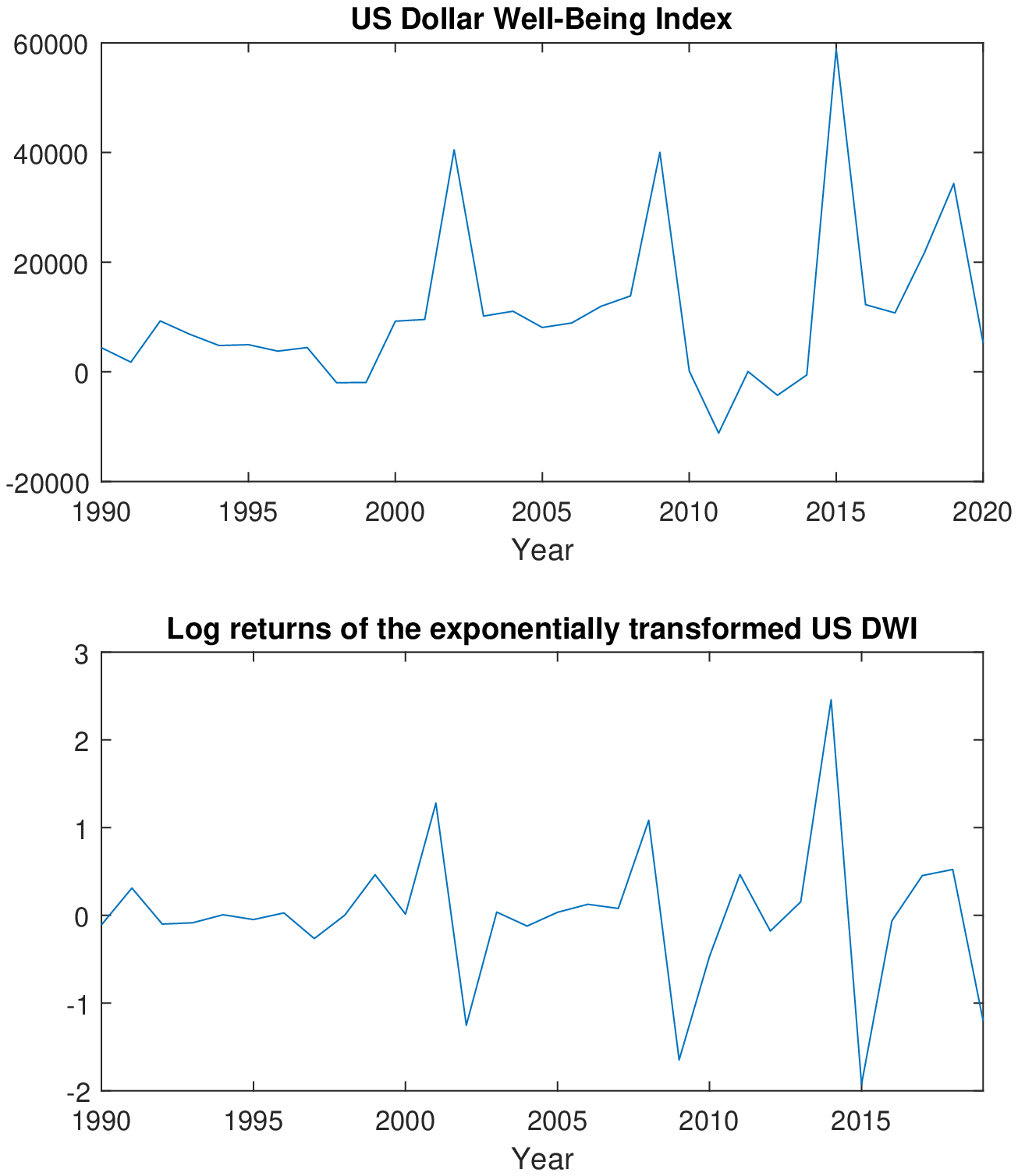}}
	\caption{(a) The US development indicators according to the World Bank reports \citep{Data}  between 1990 and 2020 and (b) the US dollar wellbeing index constructed using Eq. (\ref{Eq:DWI(l)}) and the log returns of exponentially transformed DWI subject to constraints in Eq. (\ref{Eq:f(x)})}	
	\label{Fig-US-indicators}			
\end{figure}

We write $F(k,l)$ for the $k$\textsuperscript{th} world development indicator ($k=1,\cdots,K=8$) for the $l$\textsuperscript{th} country ($l=1,\cdots,L=9$) at a given year
such that all indicators are strictly positive, i.e., $F(k,l)>0$ for all $k,l$
so that they positively contribute to the wellbeing index.
Therefore, we transform the Gini index and unemployment indicators to Neg\_Gini Index (100-Gini Index) and employment (100-unemployment), respectively.
The US development indicators which positively contribute to the US wellbeing index are shown in Figure \ref{Fig-US-indicators}(a).

\noindent We normalize each indicator of a country, $F(k,l)$, taking the corresponding indicator of all the countries ($F(k,l),  \;\; l=1,\cdots,L=9$) as follows:
\begin{equation} \label{Eq:FN(k,l)}
FN(k,l)=\frac{F(k,l)}{\sum_{l=1}^{L} F(k,l)}, \;\;\;\; k=1,\cdots, K,  \;\;\; l=1,\cdots,L.
\end{equation}

\begin{figure}[h!] 	
	\centering
	\subfigure[]{\includegraphics[width=0.55\textwidth]{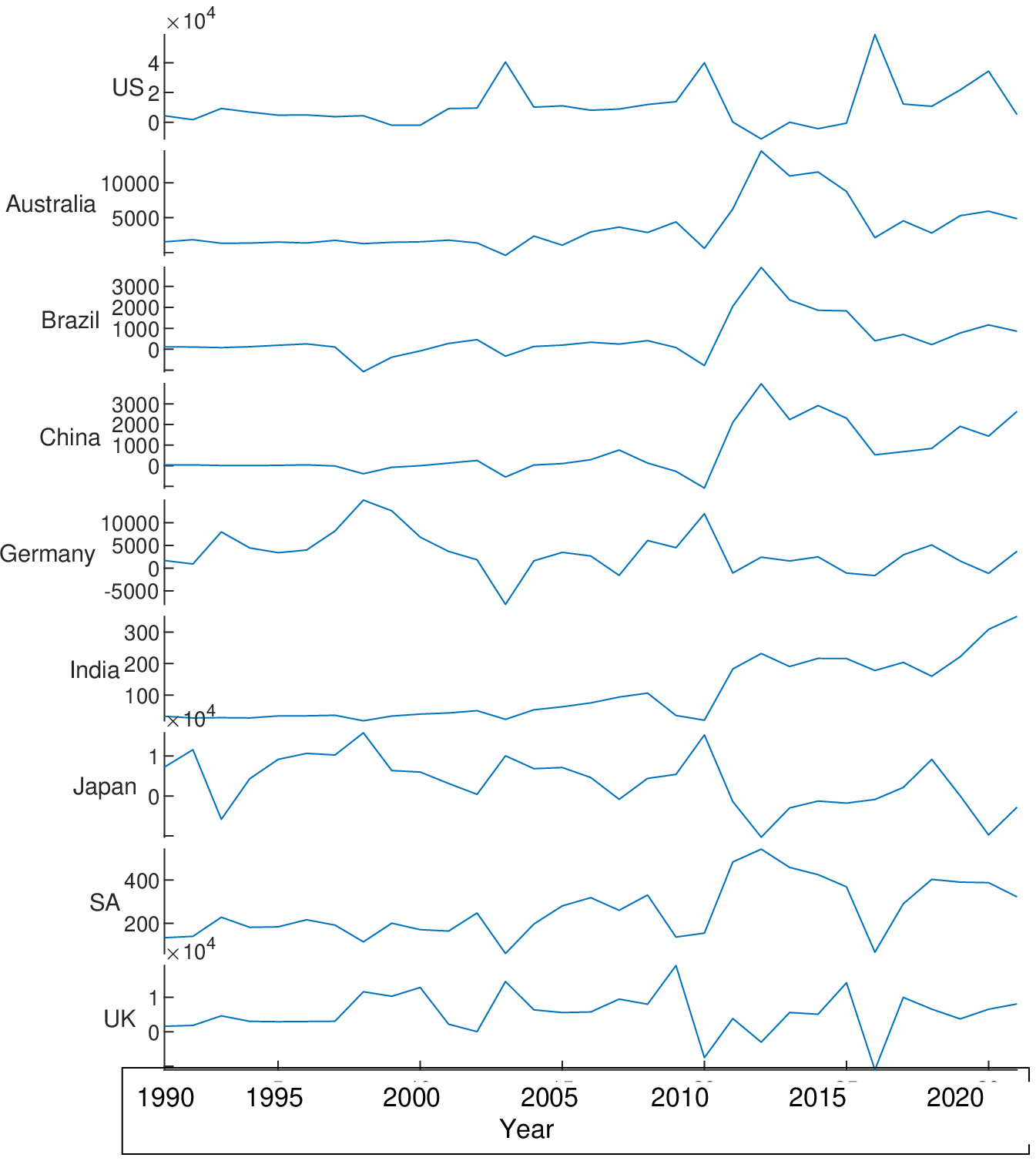}}
	\subfigure[]{\includegraphics[width=0.4\textwidth]{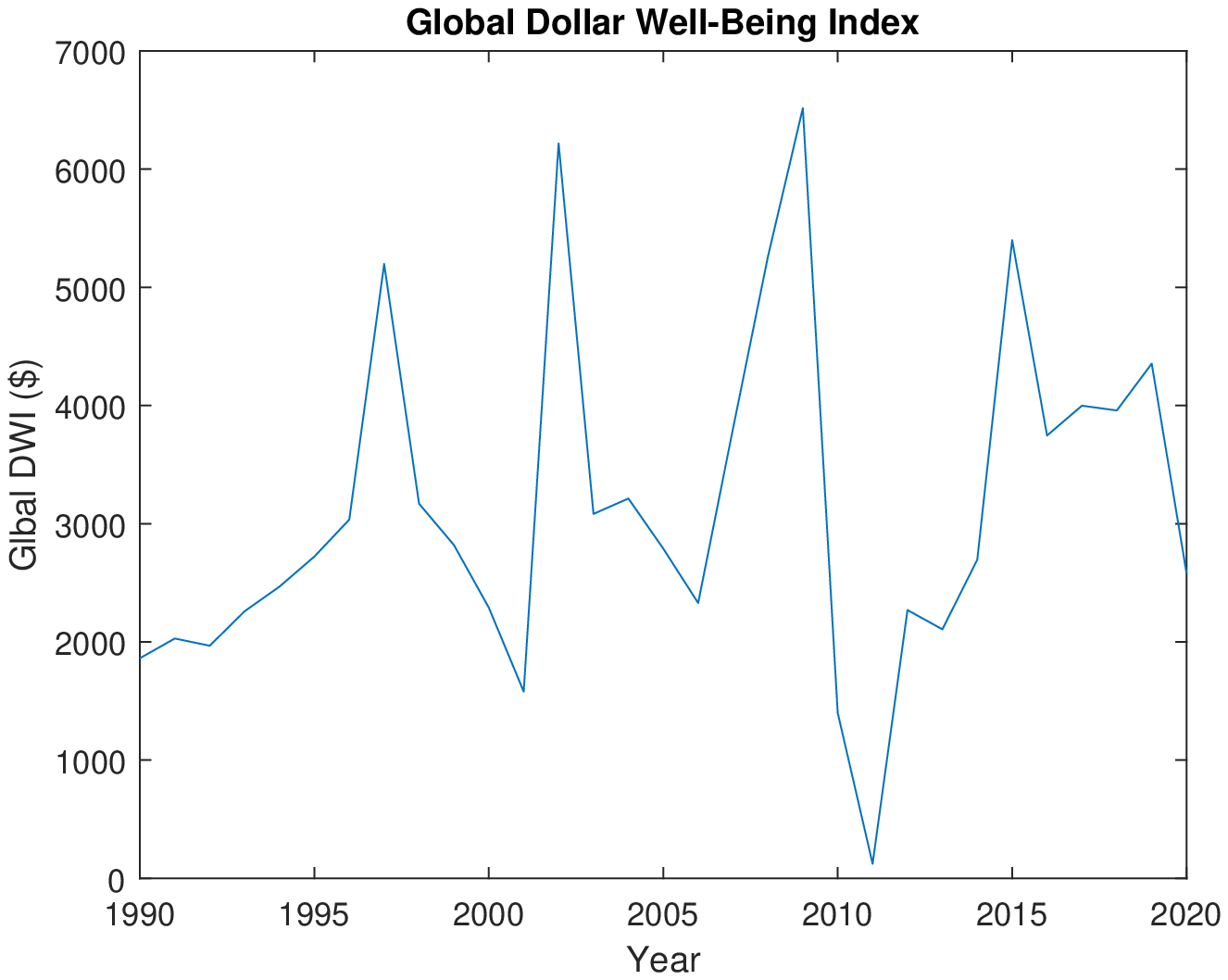}}
	\caption{(a) Dollar wellbeing indices constructed using Eq. (\ref{Eq:DWI(l)}) and (b) the global DWI proposed in Eq. (\ref{Eq:DWI})}	
	\label{Fig-DWIs}			
\end{figure}

\noindent We then define the wellbeing index for country $l$, $WI(l)$, as the average of its normalized indicators excluding GDP
\begin{equation} \label{Eq:WI(l)}
WI(l)=\frac{1}{K-1} \sum_{k=1}^{K-1} FN(k,l),   \;\;\;\; l=1,\cdots,L
\end{equation}
so that $WI(l) \in (0,1) , \; l=1,\cdots,L$. 
We monetize the US dollar value of $WI(l)$ at year $t$, $WI_t(l)$, by weighting with its corresponding GDP per capita, $GDP_t(l)$, to define a US dollar-denominated index of socioeconomic wellbeing , dollar wellbeing index (DWI), for country $l$ at year $t$ as follows:
\begin{equation} \label{Eq:DWI(l)}
DWI_t(l)=GDP_t(l) \cdot WI_t(l),  \;\;\; t=t_0=1990,\cdots,t_{30}=2020. 
\end{equation}
This indicates the ``wellbeing" of a resident of country $l$ denominated in US dollars.
For example, Figure \ref{Fig-US-indicators}(b) illustrates the wellbeing of the US per capita.
Figure \ref{Fig-DWIs}(a) shows the wellbeing of each country is different based on the dollar value per capita (DWI).
To compare the wellbeing of an individual with that of the overall population, we construct a global DWI taking the average of the DWIs:
\begin{equation} \label{Eq:DWI}
DWI_t=\frac{1}{L} \sum_{l=1}^{L} DWI_t(l),  \;\;\; t=t_0,\cdots,t_{30}.
\end{equation}
The global DWI is shown in Figure \ref{Fig-DWIs}(b).
The yearly deviations in terms of (thousands of) dollars represent the change of DWI per person from one year to another.

\subsection{Financial and econometric modeling of wellbeing indices} \label{sec:TS}

In order to model the population's wellbeing within the framework of a financial market, we must place the US dollar amount expressing the wellbeing of each representative inhabitant of a given country. 
We should view this wellbeing as the price of an asset. 
This requires a dynamic asset pricing theory approach to modeling the financial markets of these indices  of socioeconomic wellbeing, based on their observed time series. 
This approach requires the following exponential transformation of the time series of the wellbeing indices. 
We choose the exponential transformation to mimic the dynamics of asset prices, so that we can apply dynamic asset pricing theory \citep{duffie2010dynamic, schoutens2003levy}.
We determine the exponential transformation for each DWI considering the years $t=t_0,\cdots,2020$ as follows:
\begin{equation} \label{Eq:f(x)}
\begin{split}
f(x) &= a\exp(bx), \;\;\;\; a>0, b>0\\
f\left(\min_{l,t} \;\; DWI_t(l)\right)&=0 \\
f\left(\max_{l,t} DWI_t(l)\right)&=1. \\
\end{split}
\end{equation}	
Under this optimization, we set the exponential transformation of the lowest DWI to 0 and the highest DWI to 1.
We assume that the ``asset price" (the happiness of the representative inhabitant) should have a minimal value of 0 and maximal value of 1, representing 100\%. This scale is the one used in environmental, social and governance (ESG) rankings \citep{scatigna2021achievements}.
The exponential transformation for the US DWI produces $a=0.0037$ and $b=0.0001$. 
Having all DWIs positive, we define the log-returns for the exponentially transformed DWI and thus introduce the wellbeing asset pricing model for each country with respect to the global index. 

We take the log returns of the exponentially transformed DWIs as is done in dynamic asset pricing theory \citep{duffie2010dynamic}:
\begin{equation} \label{Eq:logReturns}
R_t(l)=\log \frac{f(DWI_t(l))}{f(DWI_{t-1}(l))}; \;\;\;\; l=1,\cdots,10, \;\;\; t=t_0, \cdots ,t_{30},
\end{equation}
where $f(DWI_t(l)), \;\; l=1,\cdots,9$ is the exponentially transformed DWI for the $l$\textsuperscript{th} \nc country at year $t$, and
we set $l=10$ for the exponentially transformed global DWI. 
In each country, we model the log returns of the exponentially transformed DWI with an autoregressive AR(1):
\begin{equation} \label{Eq:R_t(l)}
R_t(l) = \phi_{0} + z_{t} + \theta_{1} z_{t-1}, \;\;\;  t=t_0, \cdots ,t_{30}.
\end{equation}
where $z_{t}=\sigma_{t}\epsilon_{t}$, $\epsilon_{t}$ are assumed independent and identically distributed (iid) innovations, while $\phi_{0}$ and $\theta_{1}$ are new parameters to be estimated. 
We model the volatility ($\sigma_{t}$) using the best fit from among the time-varying volatility models ARCH(1), GARCH(1,1), and EGARCH(1,1).
The GARCH(1,1) model is defined as
	\begin{equation} \label{Eq:GARCH}
	\begin{split}
	\sigma_{t} &=\frac{z_{t}}{\epsilon_{t}} \\
	\sigma_{t}^{2} &=\alpha_{0}+\alpha_{1}z_{t-1}^{2}+\beta_{1}\sigma_{t-1}^{2}, \;\;\; t=t_0, \cdots ,t_{30}
	\end{split}
	\end{equation}	
where $\phi_{0}, \theta_{1}, \alpha_{0}, \alpha_{1}$ and $\beta_{1}$ are parameters to be estimated \citep{bollerslev1986generalized}. The sample innovations, $\epsilon_{t}$, are iid random variables with zero mean and unit variance \citep{tsay2005analysis}.


We model the log returns $R_{t}(l)$ in Eq. (\ref{Eq:R_t(l)}) using the following univariate models with standard normal iid innovations \citep{hamilton2020time, tsay2005analysis}: 
\begin{itemize}
	\item{Model 1: AR(1)-ARCH(1)}
	\item{Model 2: AR(1)-GARCH(1,1)}
	\item{Model 3: AR(1)-EGARCH(1,1)}
\end{itemize}

We compare the performances of the three models based on the Akaike information criterion (AIC) and the Bayesian information criterion (BIC), as shown in Table \ref{tab:AIC_BIC}.
For each country, we select the model which results in the lowest AIC and BIC among the estimated models (Models 1, 2, and 3).
For example, Model 1 outperforms in modeling the log returns of US DWI.
Combining the simulated sample innovations of each country, we get a 10-dimensional sample of 30 observations.
We employ $S=10,000$ scenarios for innovations based on a fitted 10-dimensional normal-inverse Gaussian distribution (NIG) which is in the domain of attraction of a 10-dimensional multivariate Gaussian distribution \citep{oigaard2004multivariate}.
By passing from normal to NIG innovations, we preserve the asymptotic unbiasedness of the parameters in the marginal time series models (\ref{Eq:GARCH}). To assess the confidence bound for those parameters we would need boostrapping methods which are beyond the scope of this paper. 
Next, using the estimated parameters in Models 1, 2, and 3 for each of the ten marginal (one-dimensional) time series, we generate $S=10,000$ Monte Carlo scenarios  of the innovations\footnote{The Basel II accord requires 10,000 scenarios in generating future portfolio returns for properly assessing the tail risk in the portfolio of returns \citep{orgeldinger2006basel, jacobson2005credit}} for the log returns in year 2021, $(R_{t_{31}}(1;s),....,R_{t_{31}}(10;s)), s=1,...,S$.
As a result, we obtain $S$ scenarios for the log returns of the ten indices for the year 2021.  As the innovations are models with NIG distribution, which captures the tail dependencies of the indices, our overall forercast of the socioeconomic market for 2021 exhibit all the ``stylized facts"  of a financial market, see \citep{rachev2000stable, taylor2011asset}. 

\begin{table}[]
\caption{Estimated dynamic model comparison for the log returns of DWIs Eq. (\ref{Eq:logReturns}) based on AIC and BIC (Model 1: AR(1)-ARCH(1), Model 2: AR(1)-GARCH(1,1), and Model 3: AR(1)-EGARCH(1,1))}
\centering
\begin{tabular}{@{}l|lll|lll|@{}}
\cmidrule(l){2-7}
                                & \multicolumn{3}{c|}{\textbf{AIC}}                                     & \multicolumn{3}{c|}{\textbf{BIC}}                                     \\ \cmidrule(l){2-7} 
                                & \multicolumn{1}{l|}{Model 1} & \multicolumn{1}{l|}{Model 2} & Model 3 & \multicolumn{1}{l|}{Model 1} & \multicolumn{1}{l|}{Model 2} & Model 3 \\ \midrule
\multicolumn{1}{|l|}{US}        & \multicolumn{1}{l|}{2.4627}  & \multicolumn{1}{l|}{2.5209}  & 2.5696  & \multicolumn{1}{l|}{2.6495}  & \multicolumn{1}{l|}{2.7544}  & 2.8498  \\ \midrule
\multicolumn{1}{|l|}{Australia} & \multicolumn{1}{l|}{1.5981}  & \multicolumn{1}{l|}{0.8630}  & 0.5255  & \multicolumn{1}{l|}{1.7849}  & \multicolumn{1}{l|}{1.0965}  & 0.8058  \\ \midrule
\multicolumn{1}{|l|}{Brazil}    & \multicolumn{1}{l|}{1.6695}  & \multicolumn{1}{l|}{1.7362}  & 1.4492  & \multicolumn{1}{l|}{1.8563}  & \multicolumn{1}{l|}{1.9697}  & 1.7294  \\ \midrule
\multicolumn{1}{|l|}{China}     & \multicolumn{1}{l|}{1.8168}  & \multicolumn{1}{l|}{1.4721}  & 1.4353  & \multicolumn{1}{l|}{2.0037}  & \multicolumn{1}{l|}{1.7056}  & 1.7156  \\ \midrule
\multicolumn{1}{|l|}{Germany}   & \multicolumn{1}{l|}{2.2366}  & \multicolumn{1}{l|}{2.3033}  & 2.1957  & \multicolumn{1}{l|}{2.4235}  & \multicolumn{1}{l|}{2.5368}  & 2.4760  \\ \midrule
\multicolumn{1}{|l|}{India}     & \multicolumn{1}{l|}{0.9228}  & \multicolumn{1}{l|}{0.9895}  & 0.5342  & \multicolumn{1}{l|}{1.1097}  & \multicolumn{1}{l|}{1.2230}  & 0.8144  \\ \midrule
\multicolumn{1}{|l|}{Japan}     & \multicolumn{1}{l|}{2.7870}  & \multicolumn{1}{l|}{2.8536}  & 2.4123  & \multicolumn{1}{l|}{2.9738}  & \multicolumn{1}{l|}{3.0872}  & 2.6926  \\ \midrule
\multicolumn{1}{|l|}{SA}        & \multicolumn{1}{l|}{2.1488}  & \multicolumn{1}{l|}{2.2155}  & 2.1473  & \multicolumn{1}{l|}{2.3357}  & \multicolumn{1}{l|}{2.4490}  & 2.4276  \\ \midrule
\multicolumn{1}{|l|}{UK}        & \multicolumn{1}{l|}{2.5711}  & \multicolumn{1}{l|}{2.6378}  & 2.4302  & \multicolumn{1}{l|}{2.7580}  & \multicolumn{1}{l|}{2.8713}  & 2.7105  \\ \midrule
\multicolumn{1}{|l|}{Global DWI} & \multicolumn{1}{l|}{2.4346}  & \multicolumn{1}{l|}{2.5013}  & 2.4220  & \multicolumn{1}{l|}{2.6214}  & \multicolumn{1}{l|}{2.7348}  & 2.7022  \\ \bottomrule
\end{tabular}
\label{tab:AIC_BIC}
\end{table}

We use the simulated scenarios and the estimated parameters of the estimated univariate time series model [AR(1)-ARCH(1) or AR(1)-GARCH(1,1) or AR(1)-EGARCH(1,1)] to forecast $S$ dynamic log returns for each DWI for 2021 ($R_{2021}$).
For example, the estimated model for the US DWI log returns is a AR(1)-EGARCH(1,1) (with a multivariate NIG sample innovations) model with the following parameters:

\begin{equation} \label{Eq:L&W}
\begin{split}
    R_{t}(l) &= 0.09 + z_t - 0.13 z_{t-1},\\ 
    \sigma_{t} &=\frac{z_{t}}{\epsilon_{t}}, \;\;\;  \\
    \sigma_{t}^{2} &= -2.97 + 0.95 z_{t-1}^{2} - 0.93 \sigma_{t-1}^{2}, \;\;\; t=t_0, \cdots ,t_{30},\\
\end{split}
\end{equation}	
where $\epsilon_{t}$ are iid random normal innovations.
Then, we simulate $S$ dynamic log returns for 2021, which we use as the dynamic asset prices for the US. 
In summary, we simulate dynamic log returns (for the year 2021, $t_{31}=2021$) for each country, by estimating econometric models for their historical log returns (between 2000 and 2020).
The joint dependence of the log returns of all indices is determined  by  a multivariate NIG distribution on the sample innovations.
Using historical and dynamic econometric modeling of the indices as time series, we provide an asset valuation risk analysis. 
Note that historical times series analyses cannot be used in dynamic asset pricing, in particular, in option pricing.
Our econometric models are designed to be consistent with dynamic asset pricing theory and thus allow the valuation of the indices as financial assets and pricing financial contracts, in particular, insurance instruments on the wellbeing indices.

\section{Measuring the tail risk of wellbeing indices} \label{sec:TailRisk}

In this section, we evaluate the performance of DWI using economic factors related to an adverse change in each country's index. 
We assess the performance of DWI with the eight world development indicators introduced in Section \ref{sec:DataDescription} to evaluate the impact of adverse events in each country's index on DWI.

According to the results of the Ljung--Box test,\footnote{See \cite{ljung1978measure}} the historical log returns inherit serial correlation and dependence. 
As dynamic log returns yield a stationary time series, our analysis focuses on dynamic innovations in capturing linear and nonlinear return dependencies.
First, we fit multivariate NIG models to the dynamic log returns of each country, that is, iid standardized innovations. 
Then, we generate $S=10,000$ scenarios from the multivariate NIG to perform the scenario analysis and compute the systemic risk measures. 
To assess the tail risk, we calculate three systemic risk measures derived from Value at Risk (VaR) \citep{philippe2001value, holton2012value}. 
We calculate VaR at the quantile levels of
$(1-\alpha)100\%$ with $\alpha=0.05$ and $\alpha=0.01$.
For $\alpha$-quantile level $(0<\alpha<1)$, $\text{VaR}_\alpha$ is defined as follows:
\begin{equation} \label{Eq:VaR}
\text{VaR}_\alpha(X) = -\inf \{x \in \mathbb{R}\; | \; F_X(x)>\alpha\}, \;\;\; 0<\alpha <1,
\end{equation}
where $F_X(x)$ is the cumulative distribution function of the log return $X$.

\cite{adrian2011covar} proposed a $\Delta$-Conditional Value at Risk (CoVaR) measure, representing in our setting the log return of a DWI, which shows how the VaR of the financial system changes when an institution experiences distress relative to its median state. We use Conditional Value at Risk (CVaR) \citep{rockafellar2000optimization, rockafellar2010buffered, acerbi2002expected, follmer2010convex, cheridito2008dual, adrian2011covar, girardi2013systemic} for finding the tail risk in log returns of DWI, denoted by $X$, at the $\alpha$ levels ((1-$\alpha$)100\%=95\% and (1-$\alpha$)100\%=99\%) as follows:  
\begin{equation} \label{Eq:CVaR}
\text{CVaR}_\alpha{(X)} =\frac{1}{\alpha} \int_{0}^{\alpha} VaR_{\gamma}(X) \; d \gamma.
\end{equation}

\noindent If $X$ has a probability density, then CVaR coincides with the expected tail loss (ETL) or expected shortfall (ES) defined as \citep{rachev2008advanced}

\begin{equation}\label{Eq:CVar_ES}
\text{CVaR}_{\alpha}(X) = -E\left(X \;  | X\leq -\text{VaR}_{\alpha}(X)\right).
\end{equation}

By switching from VaR to CVaR, which is a coherent risk measure,\footnote{See \cite{acerbi2002expected}.} we may take into account episodes of more severe distress, backtest \citep{acerbi2014backtesting} the CVaR, and enhance its monotonicity concerning the dependence parameter.
\cite{girardi2013systemic} improved how an institution is considered to be in financial difficulties, from being exactly at its $\text{VaR}_{\alpha}(X)$, $X= -\text{VaR}_{\alpha}(X)$, to being less than or equal to its -$\text{VaR}_{\alpha}(X)$, $X\leq -\text{VaR}_{\alpha}(X)$. Here, we refer to $Y$ as the global DWI log returns and $X$ as the log returns of the indices for each country in the alternative CoVaR defined in terms of the copulas in \cite{Mainik14}.
CoVaR at the level of $\alpha$, $\text{CoVaR}_\alpha$ (or $\xi_{\alpha}$), is defined as 
\begin{equation}\label{eq:def-CoVaR}
\xi_\alpha := \text{CoVaR}_\alpha := - F^{-1}_{Y|X\leq F^{-1}_{X}(\alpha)}
\left( \alpha \right) = -\text{VaR}_{\alpha}\left(Y|X\leq -\text{VaR}_{\alpha}(X)\right),
\end{equation}
where $F_{Y}$ and $F_{X}$ denote the cumulative distributions of $Y$ and $X$, respectively, and $F_{Y|X}$ is the cumulative conditional distribution of $Y$ given $X$.
An extension of CoVaR, Conditional Expected Shortfall (CoES) for DWI log returns \citep{Mainik14} at level $\alpha$ is given by
\begin{equation}\label{eq:def-CoES}
\text{CoES}_\alpha := -\E\left(Y|Y\leq -\xi_{\alpha}, X\leq -\text{VaR}_{\alpha}(X) \right).
\end{equation}

Conditional Expected Tail Loss (CoETL) is the average of the DWI losses when the DWI and the country indices extreme indicators are in distress \citep{ZariCOETL}.
CoETL quantifies the portfolio downside risk in the presence of systemic risk.
We define CoETL at level $\alpha$ as follows:
\begin{equation}\label{eq:def-CoETL}
\text{CoETL}_\alpha := -\E\left(Y|Y\leq -\text{VaR}_{\alpha}(Y),X\leq -\text{VaR}_{\alpha}(X)\right).
\end{equation}

We compare the market risk of the portfolio of wellbeing indices by calculating systemic risk measures (VaR, CoVaR, CoES, and CoETL) of the joint densities of the global DWI and DWI of each country in  Table \ref{tab_Tail}.
These numbers address how a drastic increase in the DWI of each country impacts the DWI. 
For the statistical comparison and to calculate the disastrous losses, we select 95\% and 99\%  confidence levels.
The first column reports the empirical correlation coefficients (Pearson's $R$). 
Pearson's $R$ suggests no linear relationship between Germany and DWI. 
DWI shows high positive linear relationships with US and Japan. China, Brazil, and SA have strong negative relationships with DWI, and low Pearson's $R$ values suggest weak relationships.

The table's second and third columns show each country's DWI VaR measures, i.e., the maximum losses at 95\% and 99\% confidence levels. 
These are the losses which have a 5\%, respectively, 1\% chance that the loss will exceed the maximum threshold (that is, be greater than the VaR). 
The maximum loss occurs in the US at the 99\% confidence level for both dynamic and historical log returns.
At the 95\% confidence level, the maximum loss occurs in the US (2.88) for dynamic data while the UK shows the highest loss (3.13) for historical data. 

We quantify the average loss beyond the 5\% and 1\% confidence levels via CVaR. These numbers are the expected loss of each country's worst 1\% and 5\% scenarios. 
Here the maximum average losses at  95\% and 99\% confidence levels occur in the US. 
The CoES and CoETL address how a drastic decrease in the  DWI of each country impacts the global DWI. 
For example, at the 5\% stress level, CoES provides the expected return on DWI in the highest 5\% of each country's index.  
Since CoES$_{5\%}$ and CoETL$_{5\%}$ of all countries are very close, one may  conclude that the impact of a drastic increase in the DWI of each country to global DWI is similar. 
At the highest stress level $(1\%)$, which is the worst case, the stress on the US DWI has the greatest impact on the global DWI among the other DWIs.  
The lowest impact on the global DWI comes from Australia for dynamic log returns. 


In conclusion, Table \ref{tab_Tail}  reports the market risk using the left-tail systemic risk measures on the global DWI based with respect to DWI of each country at different stress levels ($\%5$ and $\%1$). 
These findings will help investors assess the market risk of the global DWI and help them with their portfolio optimization within the global financial market of wellbeing indices.

\begin{table}[]
\centering
\begin{tabular}{@{}|lccccccccc|@{}}
\toprule
\multicolumn{10}{|c|}{Dynamic log return left-tail  risk measures}  \\ \midrule
\multicolumn{1}{|l|}{Country}   & \multicolumn{1}{c|}{\begin{tabular}[c]{@{}c@{}}Pearson's\\ R\end{tabular}} & \multicolumn{1}{c|}{\begin{tabular}[c]{@{}c@{}}VaR\\ \%95\end{tabular}} & \multicolumn{1}{c|}{\begin{tabular}[c]{@{}c@{}}VaR\\ \%99\end{tabular}} & \multicolumn{1}{c|}{\begin{tabular}[c]{@{}c@{}}CVaR\\ \%95\end{tabular}} & \multicolumn{1}{c|}{\begin{tabular}[c]{@{}c@{}}CVaR\\ \%99\end{tabular}} & \multicolumn{1}{c|}{\begin{tabular}[c]{@{}c@{}}CoES\\ \%95\end{tabular}} & \multicolumn{1}{c|}{\begin{tabular}[c]{@{}c@{}}CoES\\ \%99\end{tabular}} & \multicolumn{1}{c|}{\begin{tabular}[c]{@{}c@{}}CoETL\\ \%95\end{tabular}} & \begin{tabular}[c]{@{}c@{}}CoETL\\ \%99\end{tabular} \\ \midrule
\multicolumn{1}{|l|}{US}        & \multicolumn{1}{c|}{0.55}                                                & \multicolumn{1}{c|}{2.88}                                             & \multicolumn{1}{c|}{2.55}                                             & \multicolumn{1}{c|}{3.56}                                               & \multicolumn{1}{c|}{3.66}                                               & \multicolumn{1}{c|}{0.43}                                              & \multicolumn{1}{c|}{0.61}                                              & \multicolumn{1}{c|}{1.54}                                               & 2.65                                               \\ \midrule
\multicolumn{1}{|l|}{Australia} & \multicolumn{1}{c|}{-0.25}                                               & \multicolumn{1}{c|}{0.52}                                             & \multicolumn{1}{c|}{0.39}                                             & \multicolumn{1}{c|}{0.52}                                               & \multicolumn{1}{c|}{0.66}                                               & \multicolumn{1}{c|}{0.47}                                              & \multicolumn{1}{c|}{0.65}                                              & \multicolumn{1}{c|}{0.31}                                               & 0.53                                               \\ \midrule
\multicolumn{1}{|l|}{Brazil}    & \multicolumn{1}{c|}{-0.63}                                               & \multicolumn{1}{c|}{0.96}                                             & \multicolumn{1}{c|}{0.52}                                             & \multicolumn{1}{c|}{0.70}                                               & \multicolumn{1}{c|}{1.24}                                               & \multicolumn{1}{c|}{0.47}                                              & \multicolumn{1}{c|}{0.65}                                              & \multicolumn{1}{c|}{0.59}                                               & 0.96                                               \\ \midrule
\multicolumn{1}{|l|}{China}     & \multicolumn{1}{c|}{-0.76}                                               & \multicolumn{1}{c|}{0.76}                                             & \multicolumn{1}{c|}{0.40}                                             & \multicolumn{1}{c|}{0.62}                                               & \multicolumn{1}{c|}{1.08}                                               & \multicolumn{1}{c|}{0.47}                                              & \multicolumn{1}{c|}{0.65}                                              & \multicolumn{1}{c|}{0.36}                                               & 0.76                                               \\ \midrule
\multicolumn{1}{|l|}{Germany}   & \multicolumn{1}{c|}{0.00}                                                & \multicolumn{1}{c|}{1.31}                                             & \multicolumn{1}{c|}{0.39}                                             & \multicolumn{1}{c|}{0.58}                                               & \multicolumn{1}{c|}{1.70}                                               & \multicolumn{1}{c|}{0.46}                                              & \multicolumn{1}{c|}{0.63}                                              & \multicolumn{1}{c|}{0.55}                                               & 1.22                                               \\ \midrule
\multicolumn{1}{|l|}{India}     & \multicolumn{1}{c|}{-0.41}                                               & \multicolumn{1}{c|}{2.11}                                             & \multicolumn{1}{c|}{1.10}                                             & \multicolumn{1}{c|}{1.39}                                               & \multicolumn{1}{c|}{2.89}                                               & \multicolumn{1}{c|}{0.47}                                              & \multicolumn{1}{c|}{0.65}                                              & \multicolumn{1}{c|}{1.06}                                               & 2.12                                               \\ \midrule
\multicolumn{1}{|l|}{Japan}     & \multicolumn{1}{c|}{0.40}                                                & \multicolumn{1}{c|}{0.60}                                             & \multicolumn{1}{c|}{0.26}                                             & \multicolumn{1}{c|}{0.47}                                               & \multicolumn{1}{c|}{0.75}                                               & \multicolumn{1}{c|}{0.45}                                              & \multicolumn{1}{c|}{0.63}                                              & \multicolumn{1}{c|}{0.28}                                               & 0.58                                               \\ \midrule
\multicolumn{1}{|l|}{SA}        & \multicolumn{1}{c|}{-0.58}                                               & \multicolumn{1}{c|}{1.53}                                             & \multicolumn{1}{c|}{0.84}                                             & \multicolumn{1}{c|}{1.11}                                               & \multicolumn{1}{c|}{2.00}                                               & \multicolumn{1}{c|}{0.47}                                              & \multicolumn{1}{c|}{0.65}                                              & \multicolumn{1}{c|}{0.89}                                               & 1.54                                               \\ \midrule
\multicolumn{1}{|l|}{UK}        & \multicolumn{1}{c|}{0.24}                                                & \multicolumn{1}{c|}{0.67}                                             & \multicolumn{1}{c|}{0.37}                                             & \multicolumn{1}{c|}{0.47}                                               & \multicolumn{1}{c|}{0.80}                                               & \multicolumn{1}{c|}{0.45}                                              & \multicolumn{1}{c|}{0.64}                                              & \multicolumn{1}{c|}{0.43}                                               & 0.66                                               \\ \midrule
\multicolumn{10}{|c|}{Historical log return left-tail  risk measures}                                                                                                              \\ \midrule
\multicolumn{1}{|l|}{US}        & \multicolumn{1}{c|}{0.79}                                                & \multicolumn{1}{c|}{2.74}                                             & \multicolumn{1}{c|}{4.87}                                             & \multicolumn{1}{c|}{4.30}                                               & \multicolumn{1}{c|}{5.66}                                               & \multicolumn{1}{c|}{1.85}                                              & \multicolumn{1}{c|}{3.83}                                              & \multicolumn{1}{c|}{1.16}                                               & 4.77                                               \\ \midrule
\multicolumn{1}{|l|}{Australia} & \multicolumn{1}{c|}{-0.65}                                               & \multicolumn{1}{c|}{1.81}                                             & \multicolumn{1}{c|}{3.14}                                             & \multicolumn{1}{c|}{2.91}                                               & \multicolumn{1}{c|}{3.46}                                               & \multicolumn{1}{c|}{2.44}                                              & \multicolumn{1}{c|}{3.97}                                              & \multicolumn{1}{c|}{1.93}                                               & 3.15                                               \\ \midrule
\multicolumn{1}{|l|}{Brazil}    & \multicolumn{1}{c|}{-0.79}                                               & \multicolumn{1}{c|}{1.81}                                             & \multicolumn{1}{c|}{3.47}                                             & \multicolumn{1}{c|}{3.19}                                               & \multicolumn{1}{c|}{3.86}                                               & \multicolumn{1}{c|}{2.44}                                              & \multicolumn{1}{c|}{3.97}                                              & \multicolumn{1}{c|}{1.97}                                               & 3.49                                               \\ \midrule
\multicolumn{1}{|l|}{China}     & \multicolumn{1}{c|}{-0.82}                                               & \multicolumn{1}{c|}{2.09}                                             & \multicolumn{1}{c|}{3.73}                                             & \multicolumn{1}{c|}{3.36}                                               & \multicolumn{1}{c|}{4.23}                                               & \multicolumn{1}{c|}{2.44}                                              & \multicolumn{1}{c|}{3.97}                                              & \multicolumn{1}{c|}{2.18}                                               & 3.75                                               \\ \midrule
\multicolumn{1}{|l|}{Germany}   & \multicolumn{1}{c|}{0.04}                                                & \multicolumn{1}{c|}{2.35}                                             & \multicolumn{1}{c|}{2.80}                                             & \multicolumn{1}{c|}{2.67}                                               & \multicolumn{1}{c|}{2.97}                                               & \multicolumn{1}{c|}{2.44}                                              & \multicolumn{1}{c|}{3.97}                                              & \multicolumn{1}{c|}{2.35}                                               & 2.80                                               \\ \midrule
\multicolumn{1}{|l|}{India}     & \multicolumn{1}{c|}{-0.72}                                               & \multicolumn{1}{c|}{1.42}                                             & \multicolumn{1}{c|}{2.64}                                             & \multicolumn{1}{c|}{2.34}                                               & \multicolumn{1}{c|}{3.05}                                               & \multicolumn{1}{c|}{2.44}                                              & \multicolumn{1}{c|}{3.97}                                              & \multicolumn{1}{c|}{1.47}                                               & 2.65                                               \\ \midrule
\multicolumn{1}{|l|}{Japan}     & \multicolumn{1}{c|}{0.57}                                                & \multicolumn{1}{c|}{2.77}                                             & \multicolumn{1}{c|}{2.85}                                             & \multicolumn{1}{c|}{2.83}                                               & \multicolumn{1}{c|}{2.87}                                               & \multicolumn{1}{c|}{2.44}                                              & \multicolumn{1}{c|}{3.97}                                              & \multicolumn{1}{c|}{2.54}                                               & 2.85                                               \\ \midrule
\multicolumn{1}{|l|}{SA}        & \multicolumn{1}{c|}{-0.77}                                               & \multicolumn{1}{c|}{2.28}                                             & \multicolumn{1}{c|}{3.69}                                             & \multicolumn{1}{c|}{3.42}                                               & \multicolumn{1}{c|}{4.06}                                               & \multicolumn{1}{c|}{2.44}                                              & \multicolumn{1}{c|}{3.97}                                              & \multicolumn{1}{c|}{2.39}                                               & 3.70                                               \\ \midrule
\multicolumn{1}{|l|}{UK}        & \multicolumn{1}{c|}{-0.06}                                               & \multicolumn{1}{c|}{3.13}                                             & \multicolumn{1}{c|}{4.57}                                             & \multicolumn{1}{c|}{4.24}                                               & \multicolumn{1}{c|}{5.02}                                               & \multicolumn{1}{c|}{2.09}                                              & \multicolumn{1}{c|}{3.97}                                              & \multicolumn{1}{c|}{2.70}                                               & 4.37                                               \\ \bottomrule
\end{tabular}
\caption{The comparison of Pearson's $R$ and left-tail systemic risk measures (VaR, CoVaR, CoES, and CoETL) of the joint densities of the global DWI and DWI of each country at different stress levels for dynamic and historical log returns.}
\label{tab_Tail}
\end{table}

\section{Regression and Jensen's alphas of the wellbeing indices with respect to the global wellbeing index} \label{sec:RR}


This section describes estimating the relationships between the DWI of a country and the global DWI.
We use a regression analysis to capture the pairwise linear dependences between each DWI and global DWI as follows:
\begin{equation} \label{Eq:RR}
	Y_l = a_l + b_l Y + e_l,
\end{equation}	
where $Y_l$ and $Y$ are the log returns of the exponentially transformed DWI of country $l$ and the global index, respectively.
The terms $a_l$, $b_l$, and $e_l$ denote the intercept, gradient, and random error corresponding to the regression line, respectively.

The method of ordinary least squares (OLS) is commonly used to estimate the parameters $a_l$ and $b_l$ when the errors are normally distributed.
However, this normality assumption does not work in the presence of outliers and high leverage (influential) observations.
The method of robust regression (RR) diminishes the influence of unusual observations by iteratively reweighting least squares to assign an optimal weight to each data point \citep{knez1997robustness,hu2019modelling}.
OLS is used for short term predictions as it excludes outliers, and RR is used for long term forecasts in general.

We perform RR and OLS for the log returns of each DWI and the global DWI.
For example, we compare the performances of regression analysis for the US DWI using 
the observed data between 1990 and 2020 (historical regression) and 
the data generated for 2021 using timeseries modeling in Section \ref{sec:TS} (dynamic regression).
That is, this is regressing over the corresponding $S=10,000$ Monte Carlo scenarios for the individual index and the global index in their joint dependence \footnote{The historical regression assumes that the sample of dependent variables are iid, however, as it is clear from our econometric analysis that the dependent variables form time series that are far from being white noise. 
The log returns of the wellbeing indices exhibit heavy tailed marginal distributions and volatility clustering. That is why it is imperative to use a time series forecast as we do above and perform OLS and RR regressions on the sample of $S=10,000$ Monte Carlo iid scenarios.}.
In Table \ref{Fig-Regression_US}, we provide the estimated coefficients (intercept and gradient) for the regression models with measures of goodness of fit ($p$-values, standard errors, and root mean square error RMSE).
In dynamic RR and OLS models, the estimated coefficients are slightly similar to those with the historical models.
The dynamic regression models outperform, as they yield lower $p$-values and lower standard errors compared to the corresponding historical regression model.
Furthermore, the dynamic RR outperforms the dynamic OLS as it obtains a low RMSE and low standard errors for the estimated coefficients.

\begin{table}[]
\centering
\begin{tabular}{@{}|ll|cccc|@{}}
\toprule
\multicolumn{2}{|l|}{}                                                & \multicolumn{4}{c|}{Regression   type}                                                                                                                                    \\ \midrule
\multicolumn{1}{|l|}{\multirow{2}{*}{Data  type}} & \multirow{2}{*}{} & \multicolumn{4}{c|}{RR}                                                                                                                                                   \\ \cmidrule(l){3-6} 
\multicolumn{1}{|l|}{}                            &                   & \multicolumn{1}{c|}{Coefficient} & \multicolumn{1}{c|}{p-value} & \multicolumn{1}{c|}{\begin{tabular}[c]{@{}c@{}}Standard\\  error\end{tabular}} & RMSE                   \\ \midrule
\multicolumn{1}{|l|}{\multirow{2}{*}{Historical}} & Intercept (a)     & \multicolumn{1}{c|}{-0.026}      & \multicolumn{1}{c|}{0.705}   & \multicolumn{1}{c|}{0.070}                                                     & \multirow{2}{*}{0.381} \\ \cmidrule(lr){2-5}
\multicolumn{1}{|l|}{}                            & Gradient (b)      & \multicolumn{1}{c|}{0.673}       & \multicolumn{1}{c|}{0.000}   & \multicolumn{1}{c|}{0.093}                                                     &                        \\ \midrule
\multicolumn{1}{|l|}{\multirow{2}{*}{Dynamic}}    & Intercept (a)     & \multicolumn{1}{c|}{0.275}       & \multicolumn{1}{c|}{0.000}   & \multicolumn{1}{c|}{0.007}                                                     & \multirow{2}{*}{0.489} \\ \cmidrule(lr){2-5}
\multicolumn{1}{|l|}{}                            & Gradient (b)      & \multicolumn{1}{c|}{1.313}       & \multicolumn{1}{c|}{0.000}   & \multicolumn{1}{c|}{0.017}                                                     &                        \\ \midrule
\multicolumn{2}{|l|}{}                                                & \multicolumn{4}{c|}{OLS}                                                                                                                                                  \\ \midrule
\multicolumn{1}{|l|}{\multirow{2}{*}{Historical}} & Intercept (a)     & \multicolumn{1}{c|}{-0.008}      & \multicolumn{1}{c|}{0.937}   & \multicolumn{1}{c|}{0.093}                                                     & \multirow{2}{*}{0.512} \\ \cmidrule(lr){2-5}
\multicolumn{1}{|l|}{}                            & Gradient (b)      & \multicolumn{1}{c|}{0.862}       & \multicolumn{1}{c|}{0.000}   & \multicolumn{1}{c|}{0.125}                                                     &                        \\ \midrule
\multicolumn{1}{|l|}{\multirow{2}{*}{Dynamic}}    & Intercept (a)     & \multicolumn{1}{c|}{0.378}       & \multicolumn{1}{c|}{0.000}   & \multicolumn{1}{c|}{0.008}                                                     & \multirow{2}{*}{0.603} \\ \cmidrule(lr){2-5}
\multicolumn{1}{|l|}{}                            & Gradient (b)      & \multicolumn{1}{c|}{1.360}       & \multicolumn{1}{c|}{0.000}   & \multicolumn{1}{c|}{0.022}                                                     &                        \\ \bottomrule
\end{tabular}
\caption{Regression analysis for historical and dynamic log returns in the United States.}
\label{Fig-Regression_US}	
\end{table}

\begin{figure}[h!]
	\centering
	\includegraphics[width=0.75\textwidth]{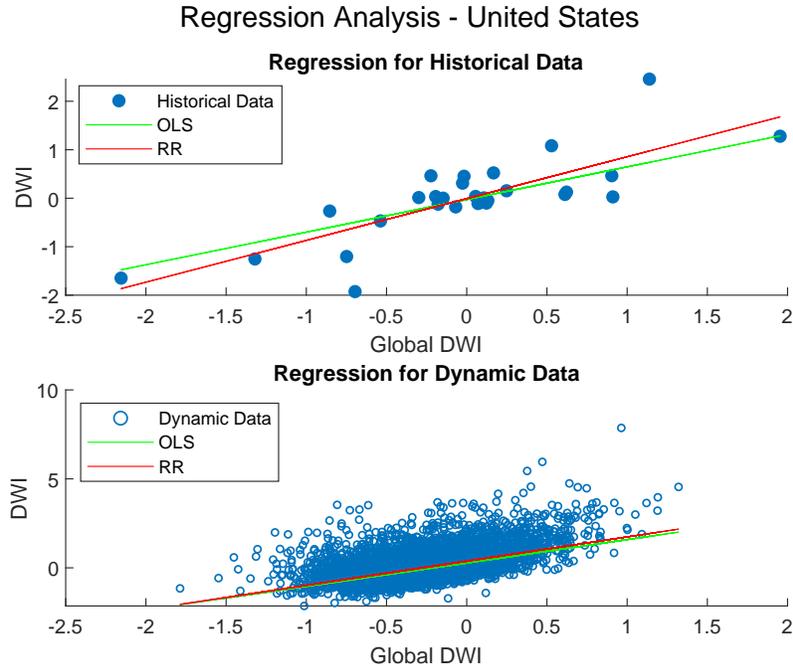} 
	\caption{Robust regression for historical and dynamic log returns in the United States. Regression lines for both historical and dynamic data result in upward forecasts.} 
	\label{Fig-DD_US}			
\end{figure}

In Figure \ref{Fig-DD_US}, we illustrate the estimated regression lines for the historical and dynamic data.
For each year in between 1990 and 2020, the historical value (data) provides an average value of the indicators we use to construct the index while 
the dynamic data are the 10,000 generated scenarios for 2021.
Historical regression fails, as the data exhibits heavy tailed asymmetry and volatility clustering.
However, the dynamic data (innovations) are in the domains of attraction of the normal distribution, see Section \ref{sec:TS}.
The dynamic model captures the AR, asymmetric volatilities, and the heavy tailed asymmetric copula (dependence).
In the presence of increasing volatility over time, the dynamic model is preferred due to the clustering of the model of volatility. 

For the US DWI, both historical and dynamic regressions result in upward forecasts, see Figure \ref{Fig-DD_US}.
The actual situation of the peoples' wellbeing around the world is grim, and this needs to be addressed.
A higher gradient of the dynamic regression indicates a better future wellbeing of the country. 
We provide the estimated gradients of the dynamic robust regressions for countries in Table \ref{Fig-RR_Slopes}.
Among the countries considered for this study, the US has the highest wellbeing while South Africa has the lowest wellbeing.
We provide a detailed regression analysis for all the countries in the appendix.

\begin{table}[]
\centering
\begin{tabular}{@{}|l|c|c|@{}}
\toprule
Country   & \begin{tabular}[c]{@{}c@{}}Estimated\\ gradient\end{tabular} & \begin{tabular}[c]{@{}c@{}}Standard\\ error\end{tabular} \\ \midrule
US        & 1.36                                                         & 0.0174                                                   \\ \midrule
Japan     & 0.11                                                         & 0.0025                                                   \\ \midrule
UK        & 0.07                                                         & 0.0031                                                   \\ \midrule
Germany   & 0.01                                                         & 0.0125                                                   \\ \midrule
Australia & -0.22                                                        & 0.0092                                                   \\ \midrule
Brazil    & -0.77                                                        & 0.0095                                                   \\ \midrule
China     & -1.04                                                        & 0.0087                                                   \\ \midrule
India     & -1.11                                                        & 0.0025                                                   \\ \midrule
SA        & -1.14                                                        & 0.0157                                                   \\ \bottomrule
\end{tabular}
\caption{Robust regression for historical and dynamic log returns in the United States. Regression lines for both historical and dynamic data result in upward forecasts.}
\label{Fig-RR_Slopes}	
\end{table}


In the time series of the log returns of the indices, negative drops are more pronounced and those downturns will be captured by the NIG distribution of the innovations \citep{schlosser2011normal}.
Therefore, we use AR(1)-ARCH(1,1) or AR(1)-EGARCH(1,1) with multivatriate NIG for option pricing, which has flexible probability distributions capturing heavy tails, central and tail dependencies. 
The dynamic models allow us to monetize the financial insurance instruments.
Historical (static) methods cannot deliver proper pricing models on wellbeing indices as they can't deliver no-arbitrage values for insurance instruments. 
Thus, we must rely on no-arbitrage asset pricing theory, and build dynamic models when pricing the wellbeing of individual countries and the world.
Hence, we use dynamic predictive models when forecasting future trends in the world wellbeing country by country and globally.

Our global index (DWI) will serve as a ``market index,” while the attached US index is an example of a ``risky asset” within the market of wellbeing indices.
Thus we want to know the “beta” in capital asset pricing model (CAPM) \citep{markovits2007organizational}, for the US index, and if more countries are involved we want to have an analogue of the Fama French three factor model \citep{fama1992cross, fama1993common} on the market of wellbeing indices. 
That is, which countries contribute most to the wellbeing of the world. \\

Jensen's alpha \citep{jensen1968performance} shows the average return on a portfolio or investment above or below the portfolio benchmark\footnote{In market equilibrium, the Jensen alpha is equal to zero.
However, as argued by George Soros, the real financial market is always in a pre-equilibrium, leading to a nonzero Jensen's alpha (equilibrium itself has rarely been observed in real life — market prices have a notorious habit of fluctuating.) \citep{soros2015alchemy}}.  We apply Jensen's alpha to determine the maximum possible return on each country's DWI and the performance of each country's DWI against the global DWI. 

Table \ref{Jensen_alpha} exhibits the value of Jensen's alpha for each country's DWI. 
The alphas of historical indices are not statistically significant, as they are close to zero, showing no evidence the country's DWIs are better than the global DWI. However, we observed different results for dynamic indices. For example, the positive alpha of 0.5359 and 0.5306 for South Africa and India, respectively, suggest that the risk-adjusted return of the DWI of these countries is much higher than the global DWI. On the other hand, the values of Jensen's alpha for Japan, Germany, and the US, which are statistically significant and negative, show that the global DWI outperforms these nations. The results show that the returns of the UK, Australia, and China compared to the global DWI is roughly similar.

\begin{table}[]
\centering
\begin{tabular}{@{}lcc@{}}
\toprule
Coutry         & Historical   Index & Dynamic Index \\ \midrule
Japan          & -0.048             & -0.5247       \\
Germany        & 0.0087             & -0.2615       \\
United States  & -0.0075            & -0.2264       \\
United Kingdom & 0.0232             & 0.1487        \\
Australia      & 0.0242             & 0.189         \\
China          & 0.0558             & 0.1951        \\
Brazil         & 0.0198             & 0.4666        \\
India          & 0.0894             & 0.5306        \\
South Africa   & 0.0404             & 0.5359        \\ \bottomrule
\end{tabular}
\caption{Jensen's Alpha CAPM for historical and dynamic indices}
\label{Jensen_alpha}
\end{table}

\section{Efficient frontier of the markets of country's wellbeing risk measures 
} \label{sec:EF}

In accordance with \cite{markowitz1952}'s method, the goal of portfolio optimization is to identify the daily set of weights $w$ that reduces the return risk of the portfolio (for that day) subject to a desired expected return $(r_p)$. The targeted return value indicates the investor's risk tolerance: the higher the value, the more risk the investor is willing to accept. We can describe the mean-variance and mean-CVaR optimization \citep{uryasev2001conditional} as the goal of minimizing the portfolio variance $\sigma_p$ and the portfolio CVaR denoted by CVaR$_{p,\alpha}$ subject to a preferred expected return by using the variance and CVaR as the risk measure.  Consider a portfolio consisting of $n$ risky assets with daily return values $r(t)=(r_1(t),r_2(t),...,r_n(t))$, having the portfolio mean and standard deviation $\bar{r} = (\bar{r_1}, \bar{r_2}, ..., \bar{r_3})$ and $\sigma_p=(\sigma_1,\sigma_2, ..., \sigma_n)$, respectively. 

\begin{equation}
    \text{minimize\,\,} w' \sigma_p w \,\, \text{subject to} \,\, \bar{r}w = r_p \,\,\text{and} \,\,\sum_{i=1}^{n} w_i = 1
\end{equation}
Because the target return varies, the best solution for $(\sigma_p, r_p)$ results in a hyperbola curve known as the portfolio frontier. The efficient frontier (EF) is the region of the portfolio frontier where projected mean returns exceed $r_p$. 
Consider a portfolio consisting of $n$ risky assets with daily return values $r_p$, with the expected risk adjusted returns mean return $E(r_p)$ and the risk measure $V(r_p)$. The portfolio optimization can be summarized as follows, 
\begin{equation}
    \text{min}_w (-\gamma E(r_p) - (1-\gamma) V(r_p)) \,\, \text{subject to}  \,\,\sum_{i=1}^{n} w_i = 1
\end{equation}
where the risk-aversion parameter $\gamma  \in [0,1]$ determines the positions along the EF, with $\gamma = 0$  corresponding to the minimum-risk portfolio.

For each country's DWI index, the optimization applied to the ensemble of the target returns produces an EF by considering the variance, CVaR$_{p,0.05}$, and CVaR$_{p,0.01}$ risk measures, to contrast the effects of central and tail risk on optimization, and since the standard deviation, which is not a coherent risk measure. The DWI dynamic and historical indices of Section \ref{sec:DWI} are used to illustrate the EFs of each country. 

\begin{figure}[h!]
	\centering
	\includegraphics[width=0.75\textwidth]{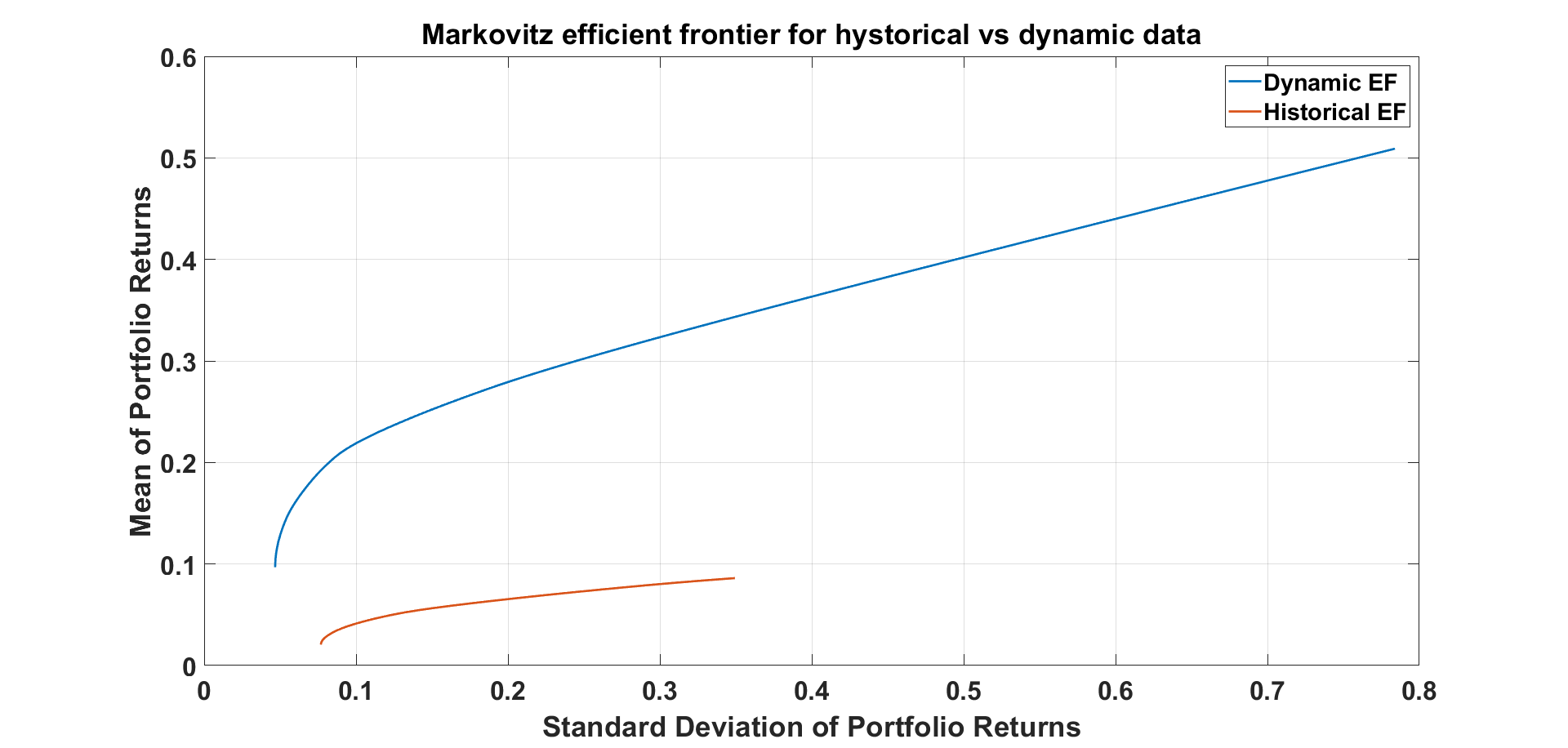} 
	\caption{Markovitz Efficient Frontier}
	\label{Fig-EF}			
\end{figure}

Figure \ref{Fig-EF} plots EFs computed for the US dynamic and historical DWI index. The mean-variance risk measure was used to generate efficient borders for the US dynamic and historical DWI index. We used the set of equally spaced values $\gamma={0,0.01,...,0.99}$ in plotting each EF. Historical EF is short since we have only 30 historical data. We expect a long EF for simulated data. Standard deviation increased in the dynamic EF from $0.05$ to $0.8$ and in the same frontier for historical data, from $0.08$ to $0.34$. The growth in $E(r p)$ of dynamic and historical DWI is consistent with the same observation; however, the increases are less evident for the EF. 

\begin{figure}[h!] 	
	\centering
	\subfigure[]{\includegraphics[width=0.45\textwidth]{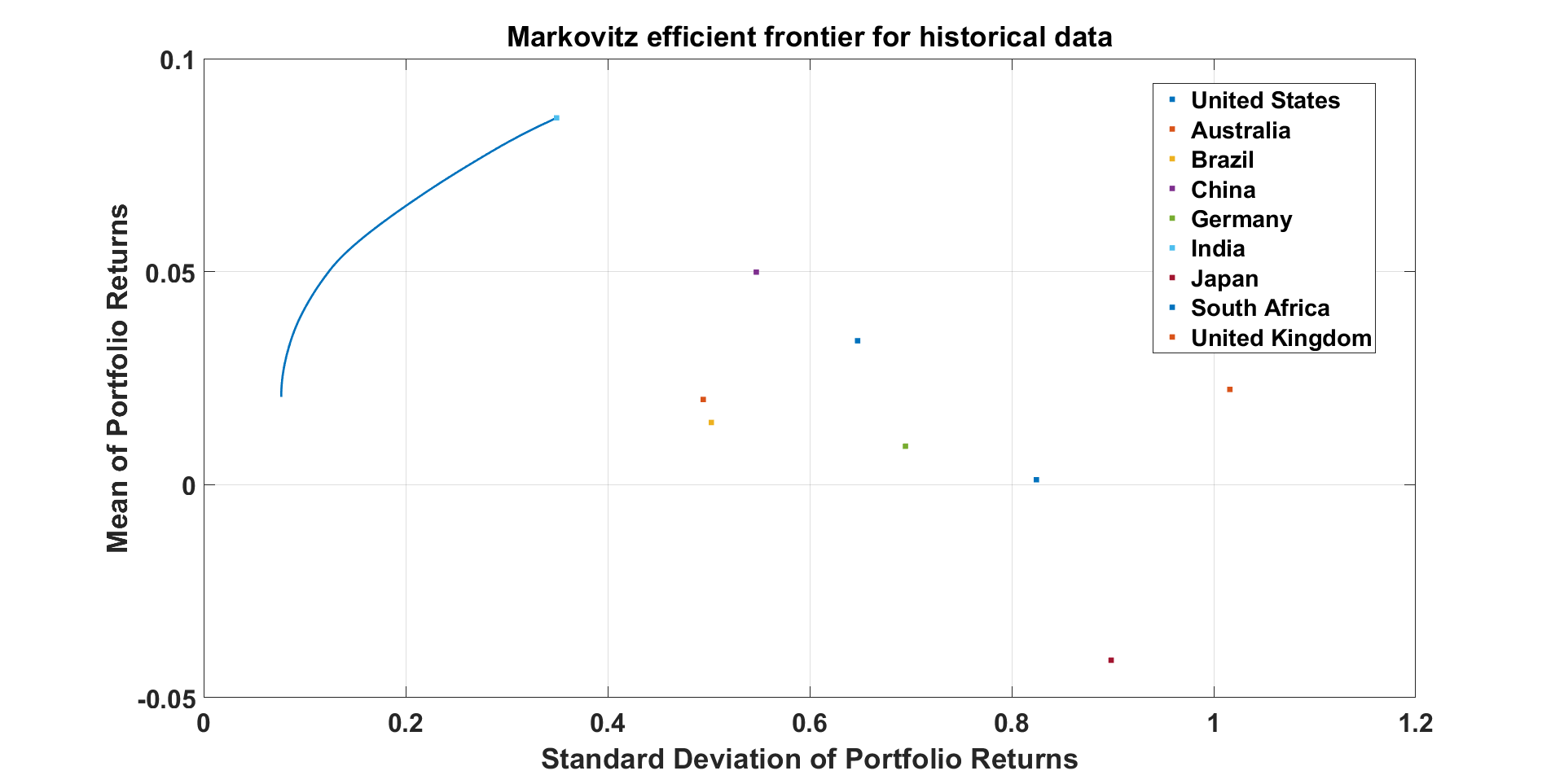}}
	\subfigure[]{\includegraphics[width=0.45\textwidth]{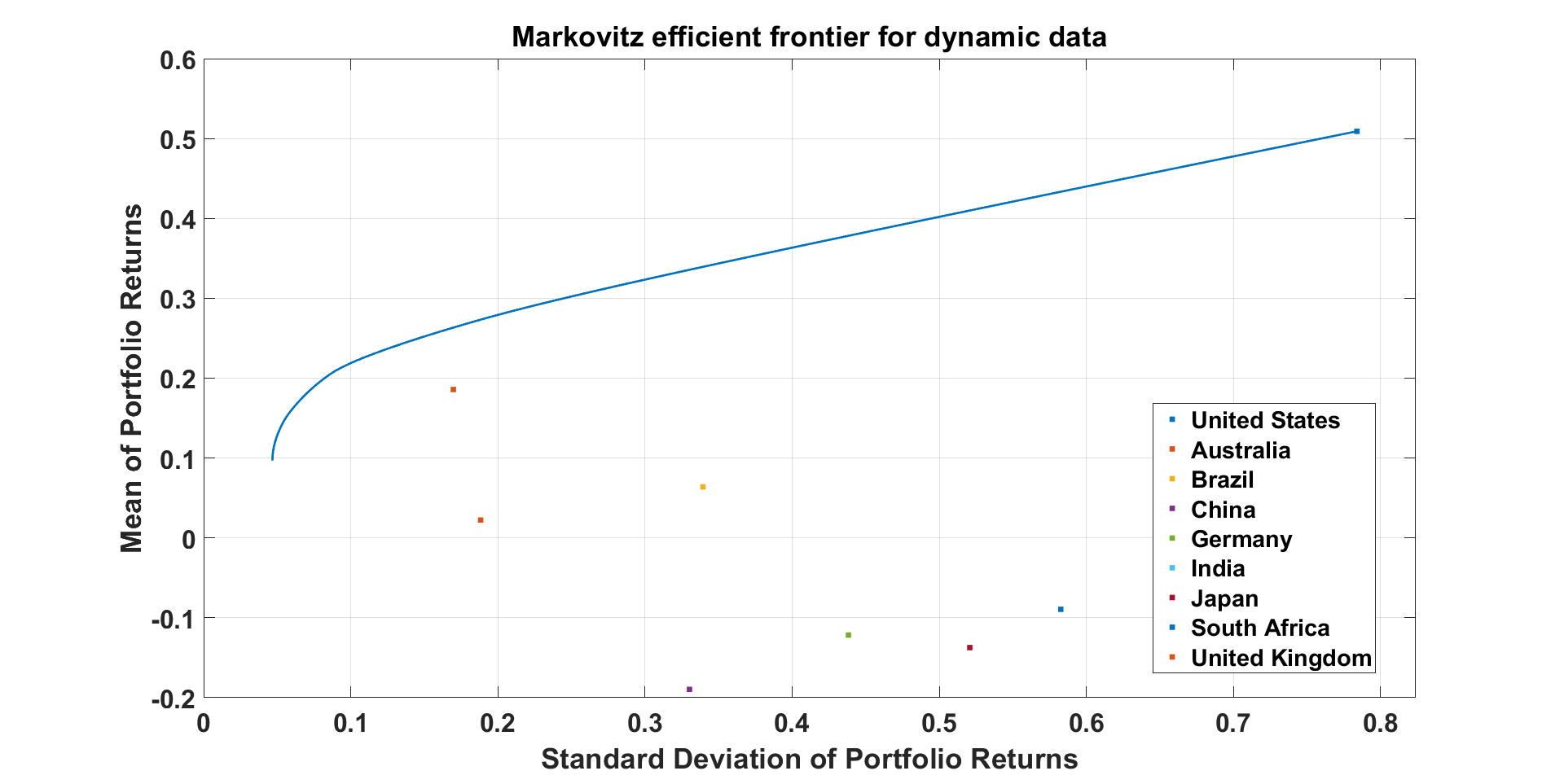}}
	\caption{Markovitz Efficient Frontier: (a) Historical Indices and (b) Dynamic Indices.}	
	\label{Fig-EF-CVaR}			
\end{figure}

Figures \ref{Fig-EF-CVaR}.a and \ref{Fig-EF-CVaR}.b reproduce Figure \ref{Fig-EF} for the tail risk measures CVaR$_{p,0.05}$, and CVaR$_{p,0.01}$. There are overall qualitative similarities in the behaviors of the CVaR EFs compared to the mean-variance EFs. The EFs change “smoothly” and are convex. However, the variation in the behavior of the EFs is more pronounced under the CVaR risk measures compared to the mean-variance EF.

\begin{figure}[h!] 	
	\centering
	\subfigure[]{\includegraphics[width=0.45\textwidth]{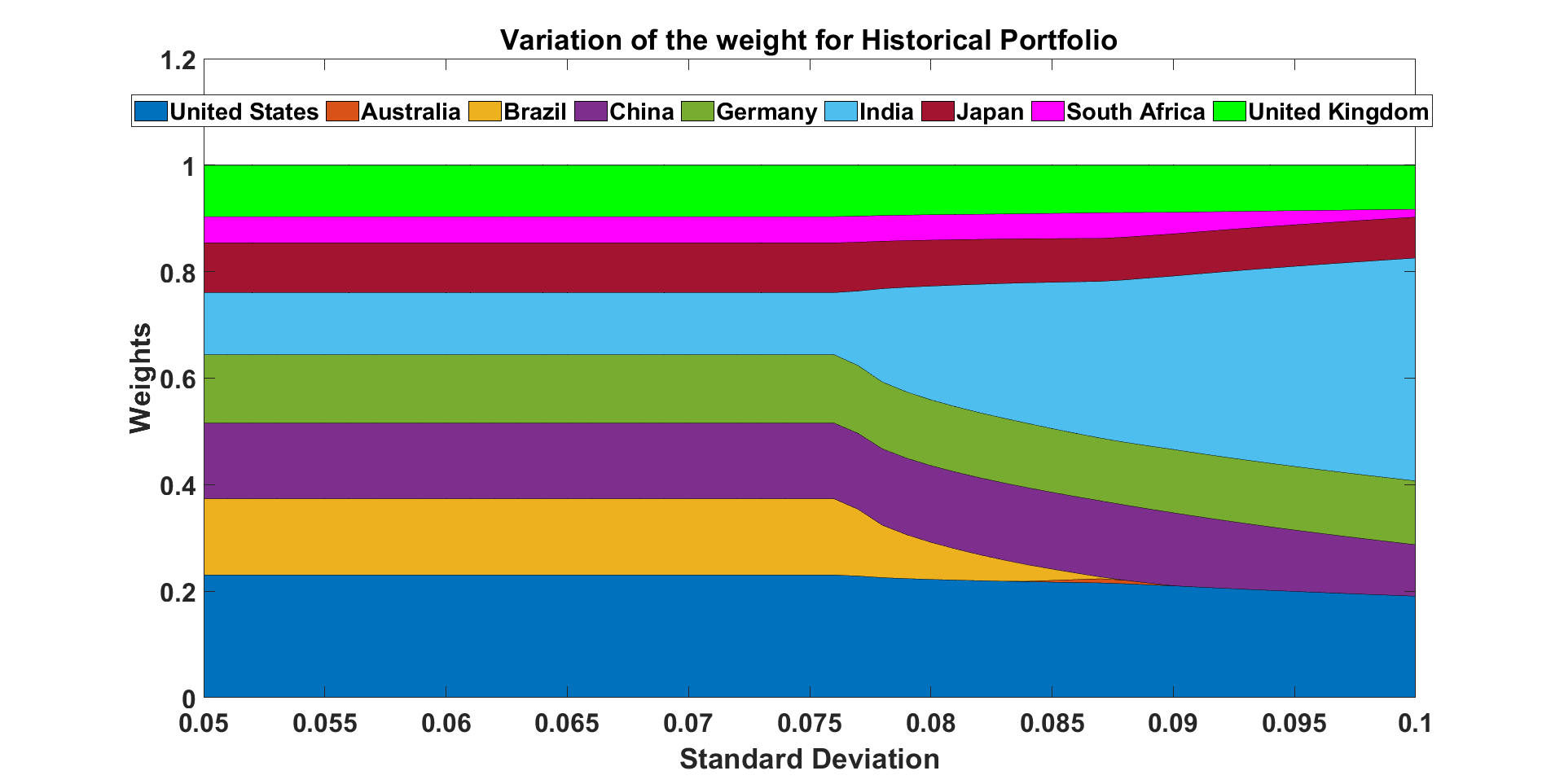}}
	\subfigure[]{\includegraphics[width=0.45\textwidth]{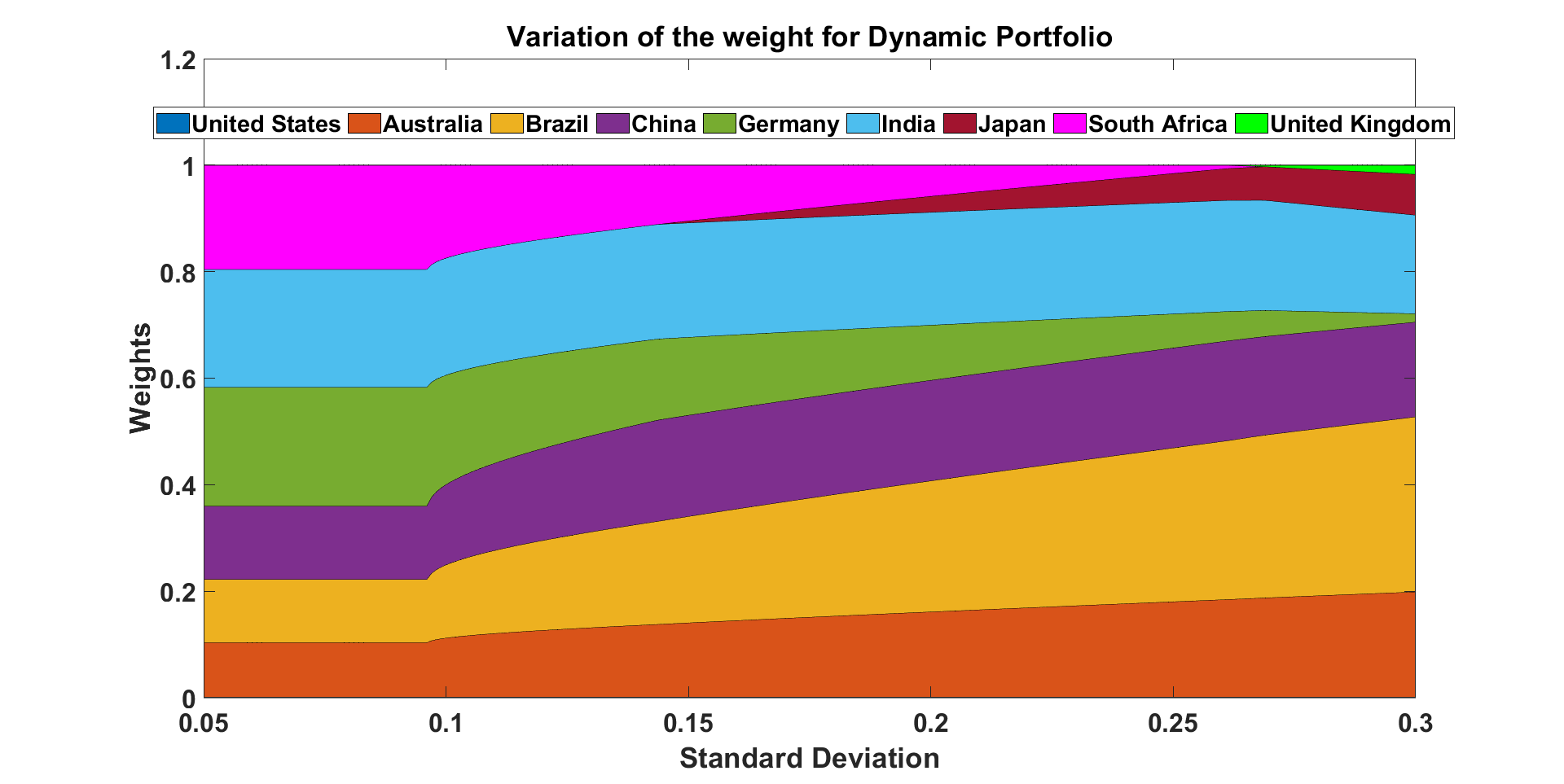}}
	\caption{Variation of the weight composition of the Markovitz optimal portfolios along each efficient frontier (as a function of standard deviation) (a) historical portfolio and (b) dynamic portfolio.}	
	\label{Fig_Weight_STD}			
\end{figure}

\begin{figure}[h!] 	
	\centering
	\subfigure[]{\includegraphics[width=0.45\textwidth]{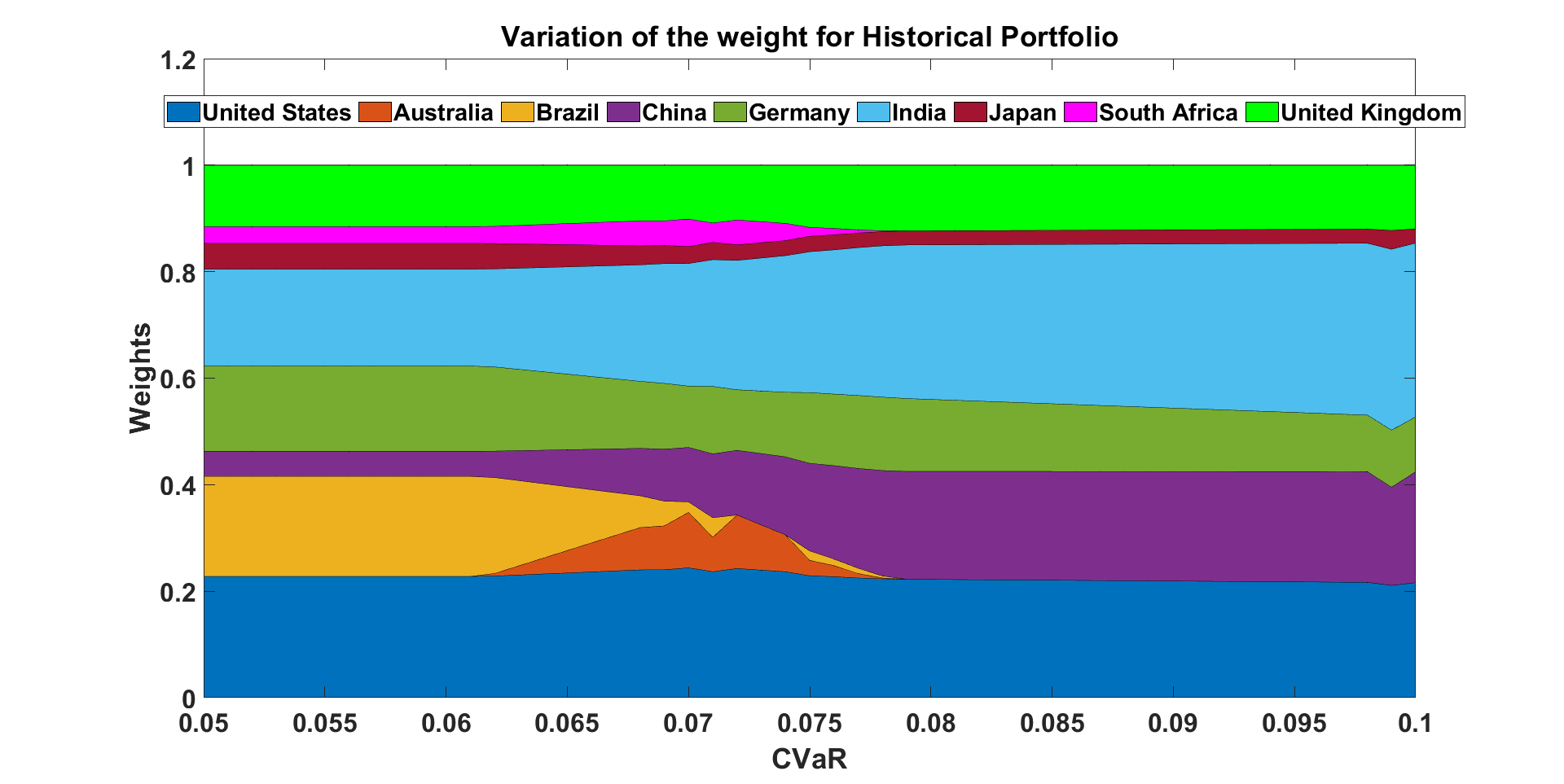}}
	\subfigure[]{\includegraphics[width=0.45\textwidth]{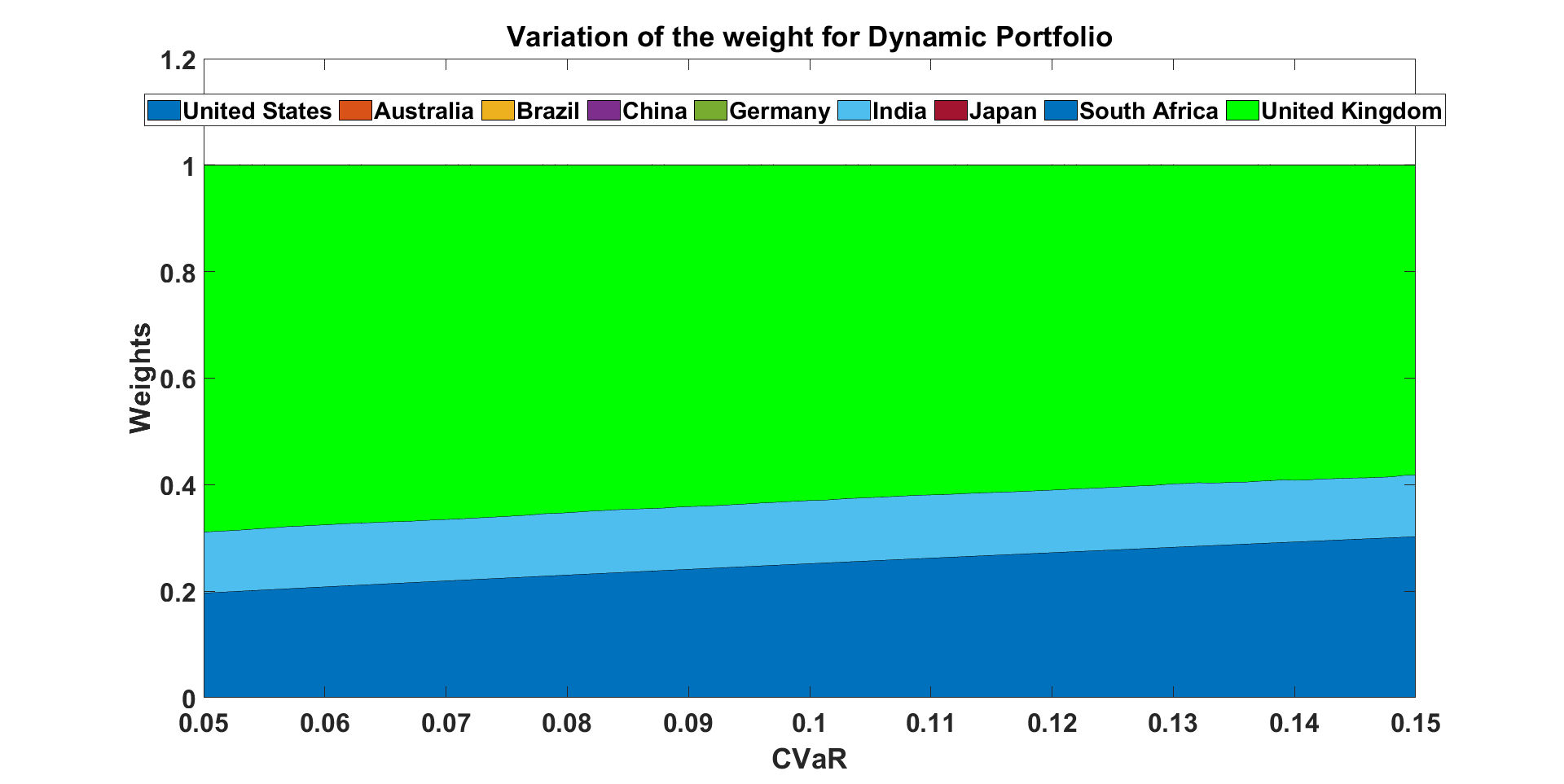}}
	\caption{Variation of the weight composition of the CVaR$_{p,\alpha}$ optimal portfolios along each efficient frontier (as a function of $\alpha$) (a) historical portfolio and (b) dynamic portfolio.}	
	\label{Fig_Weight_CVAR}			
\end{figure}

In Figures \ref{Fig_Weight_STD} and \ref{Fig_Weight_CVAR}, we depict the optimal portfolio weight composition along the EFs of the mean-variance and CVaR optimizations. For the historical portfolio EFs, the weights of the portfolios of India and China increase as the standard deviation increases. In other words, at high values of the standard deviation and CVaR, the optimizer focuses on China, India, and the United States. However, for the dynamic portfolio EFs, the weights of the portfolios of Brazil, China, and Australia increase with increasing standard deviation. Also, at high values of the standard deviation and CVaR, the optimizer focuses on the United Kingdom and South Africa. This shows that these countries are the safest in case of a drastic increase in DWI indices.

\begin{figure}[h!] 	
	\centering
	\subfigure[]{\includegraphics[width=0.45\textwidth]{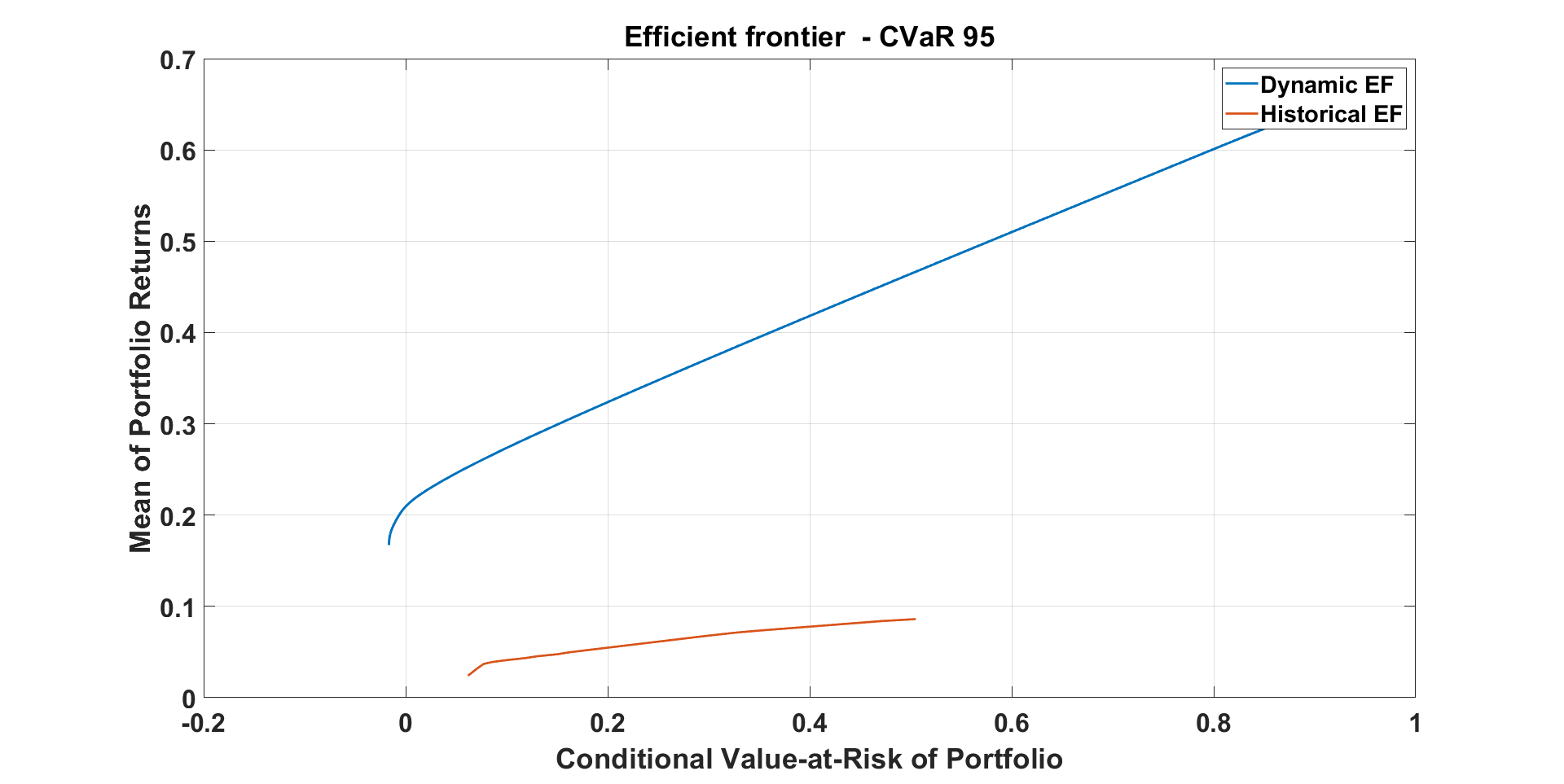}}
	\subfigure[]{\includegraphics[width=0.45\textwidth]{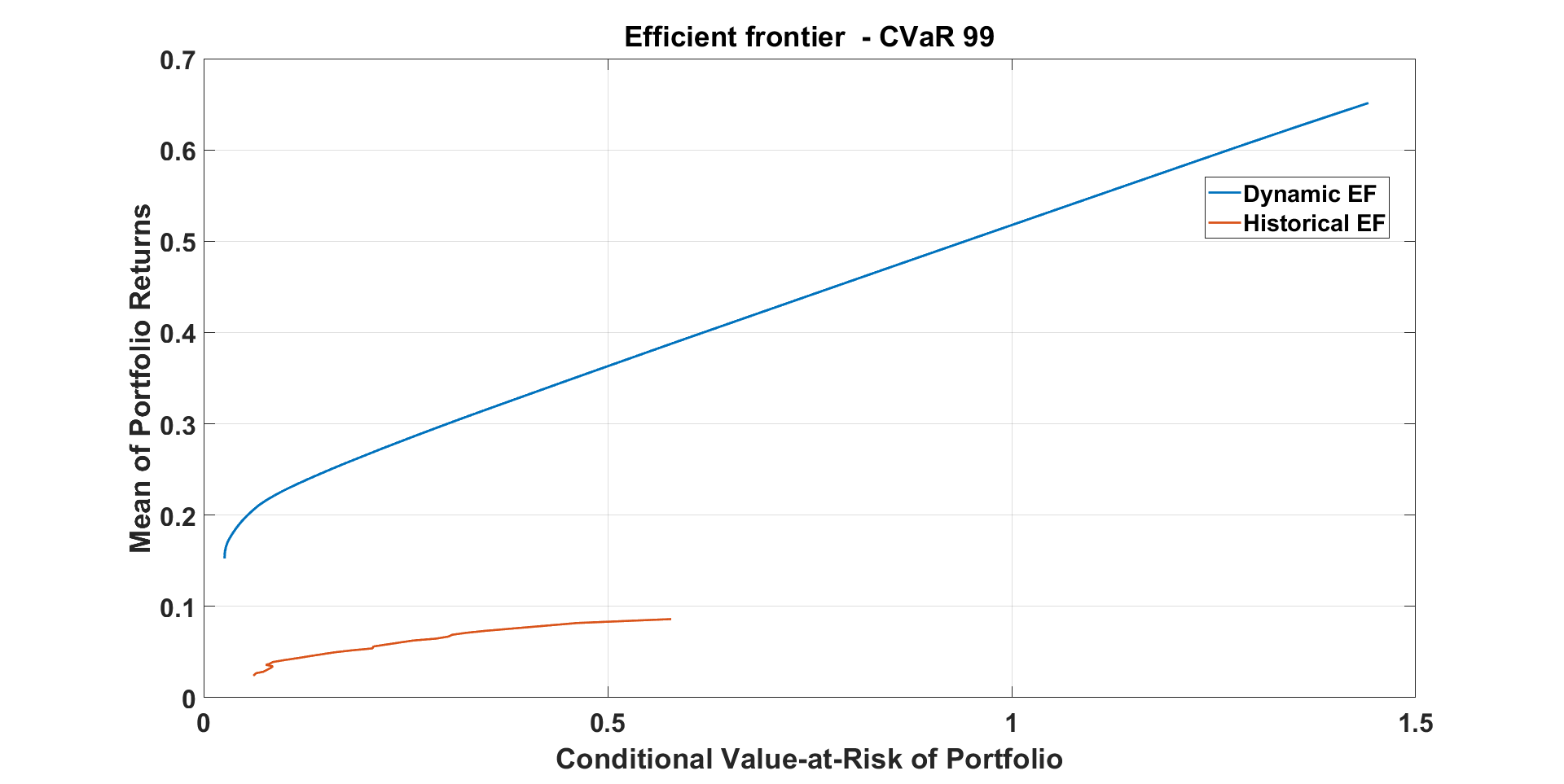}}
	\caption{Conditional value-at-risk portfolio optimization: (a) CVaR$_{p,0.05}$, and (b) CVaR$_{p,0.01}$ EFs.}	
	\label{Fig_EF_CVaR}			
\end{figure}

\subsection{Efficient frontier and risk measures of the market for countries with high GDPs}

The rankings of national economies have changed considerably over time. Nominal GDP per capita is a metric for comparing national wealth. So, it is interesting to analyze the portfolio of wellbeing indices of countries with the largest GDPs and compare them with all countries' portfolios of wellbeing indices. According to the most recent data from the World Bank, the countries with the highest nominal GDP in 2022 are the United States: \$20.89 trillion, China: \$14.72 trillion, Japan: \$5.06 trillion, and Germany: \$3.85 trillion, respectively. 

Here, we analyze a portfolio of wellbeing indices of countries with high GDPs. Just as in the last section for the four countries with the highest GPDs, the optimization applied to the ensemble of the target returns produces an EF by considering the variance, CVaR$_{p,0.05}$, and CVaR$_{p,0.01}$ risk measures, to contrast the effects of central and tail risk on optimization. A new dynamic index explained in Section \ref{sec:DWI} is created for these four countries, and these new indices are used to illustrate the EFs of each country.

\begin{figure}[h!]
	\centering
	\subfigure[]{\includegraphics[width=0.45\textwidth]{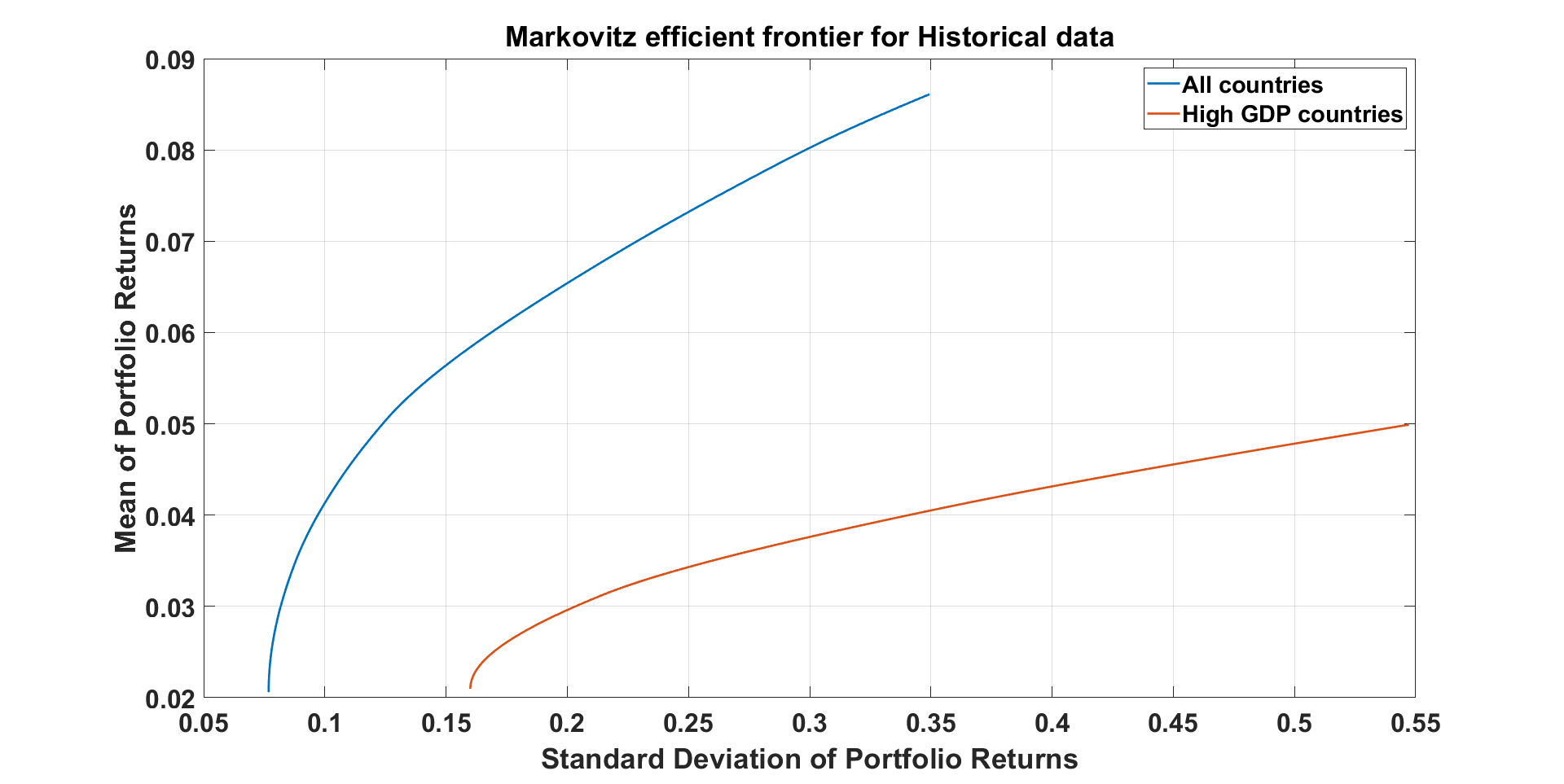}}
	\subfigure[]{\includegraphics[width=0.45\textwidth]{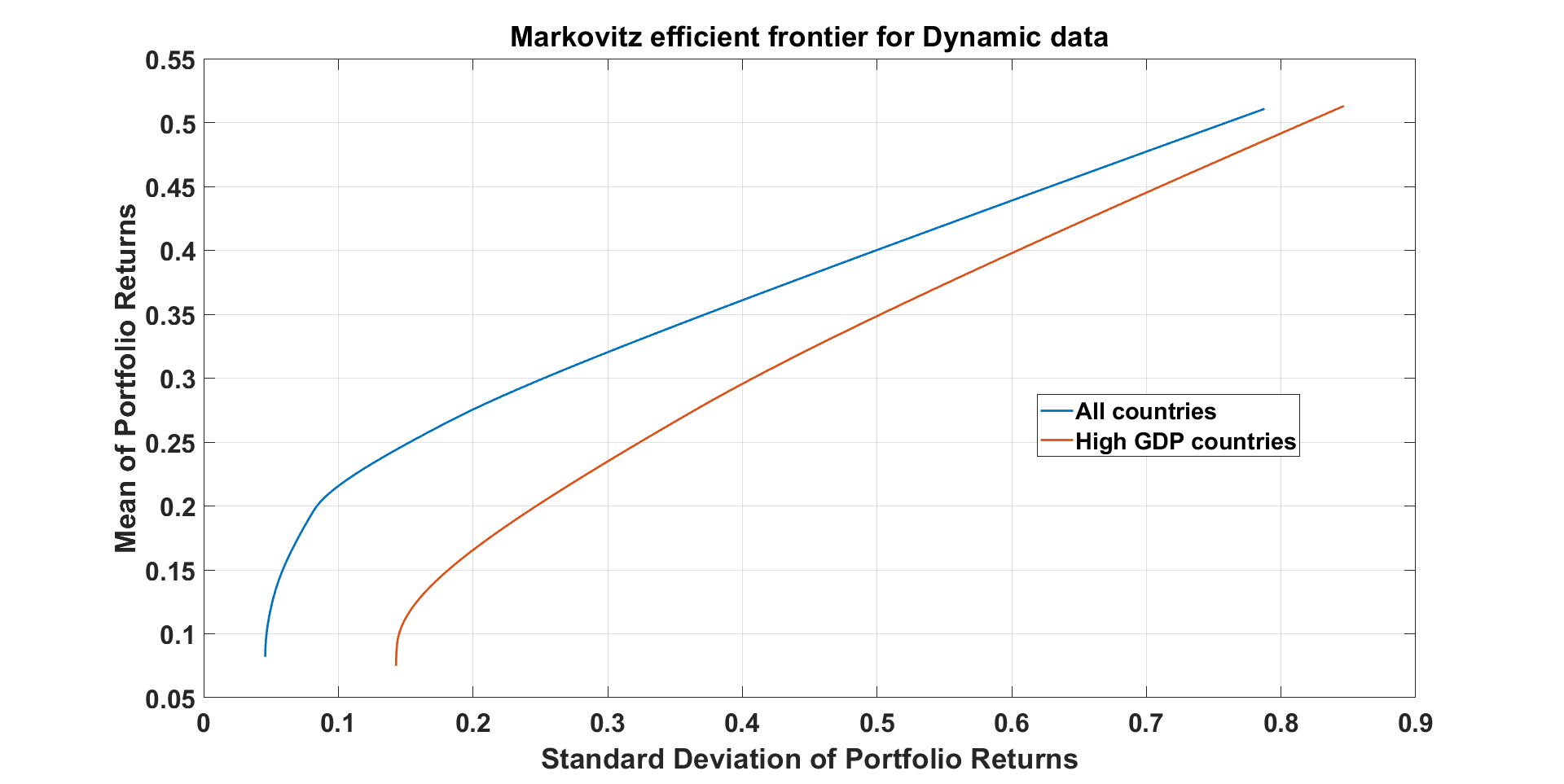}}
	\caption{Markovitz Efficient Frontier: (a) Historical Indices and (b) Dynamic Indices.}	
	\label{Fig-EF-Mark-four}			
\end{figure}

Figures \ref{Fig-EF-Mark-four}.a and \ref{Fig-EF-Mark-four}.b plot EFs computed for historical and dynamic portfolio of wellbeing
indices for the countries with the largest GDPs and compare them with the EFs of portfolio wellbeing indices for all countries. The mean-variance risk measure was used to generate efficient borders for the dynamic and historical DWI indices. As we explained in the previous section, the historical EF is short since we have only 30 historical pieces of data and expect a long EF for simulated data. The standard deviation increased in the historical EF for the high-GDP countries, from $0.16$ to $0.55$, and in the same frontier for all countries, from $0.07$ to $0.34$. 
The growth in $E(r_p)$ of DWI of all countries is higher than for the countries with high GDP at the same level of risk. Therefore, this indicates that for the same expected return value, the risk of investing in the DWI of countries with a high GDP is higher than in the DWI of all countries. Overall qualitative similarities exist in the behaviors of the dynamic, EFs for DWI of all countries and countries with high GDPs. However, for historical data, the EFs of DWI for countries with high GDPs change ``smoothly" and are convex, while intense changes are observed in the EF of DWI for all countries.

\begin{figure}[h!] 	
	\centering
	\subfigure[]{\includegraphics[width=0.45\textwidth]{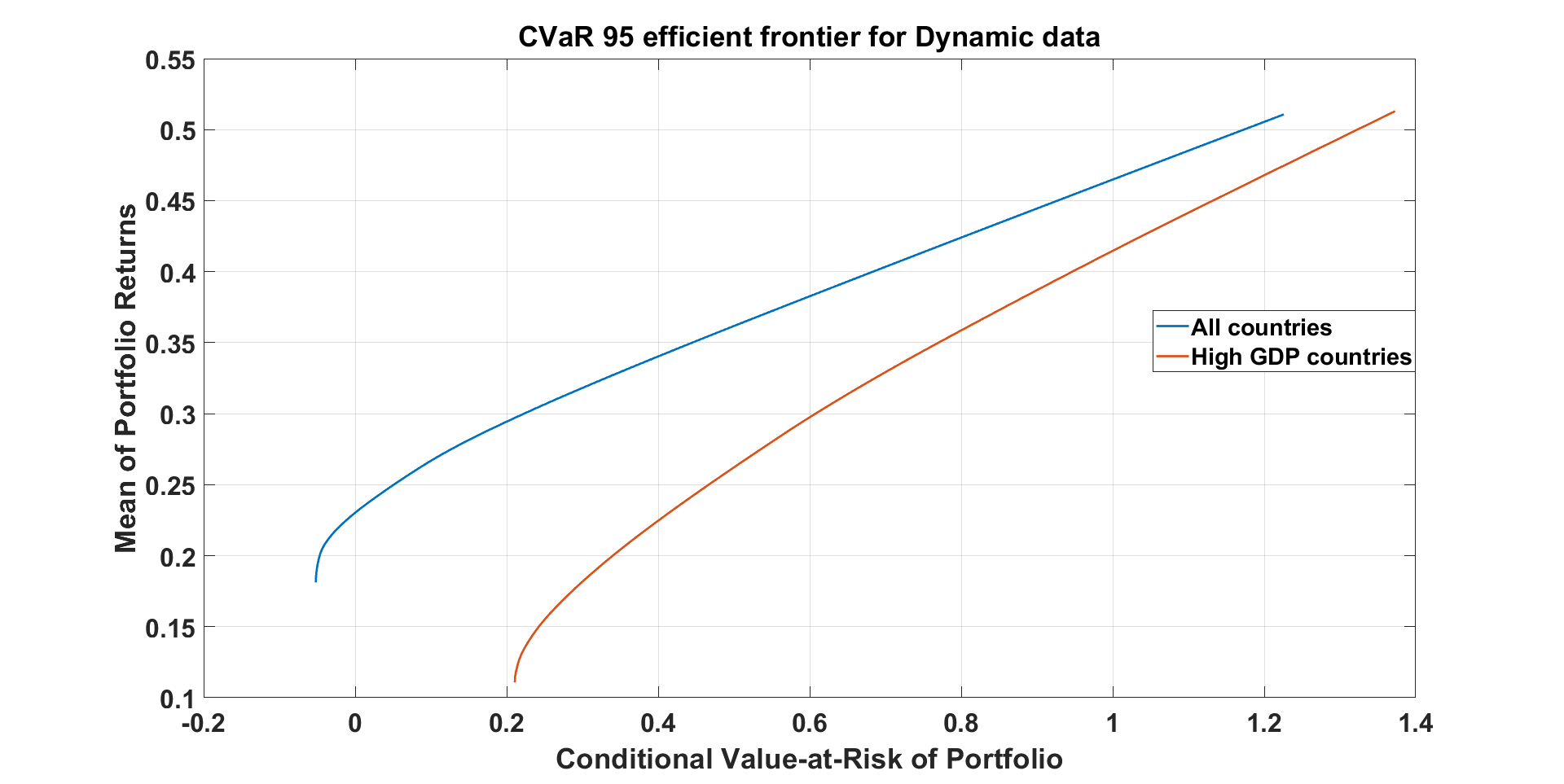}}
	\subfigure[]{\includegraphics[width=0.45\textwidth]{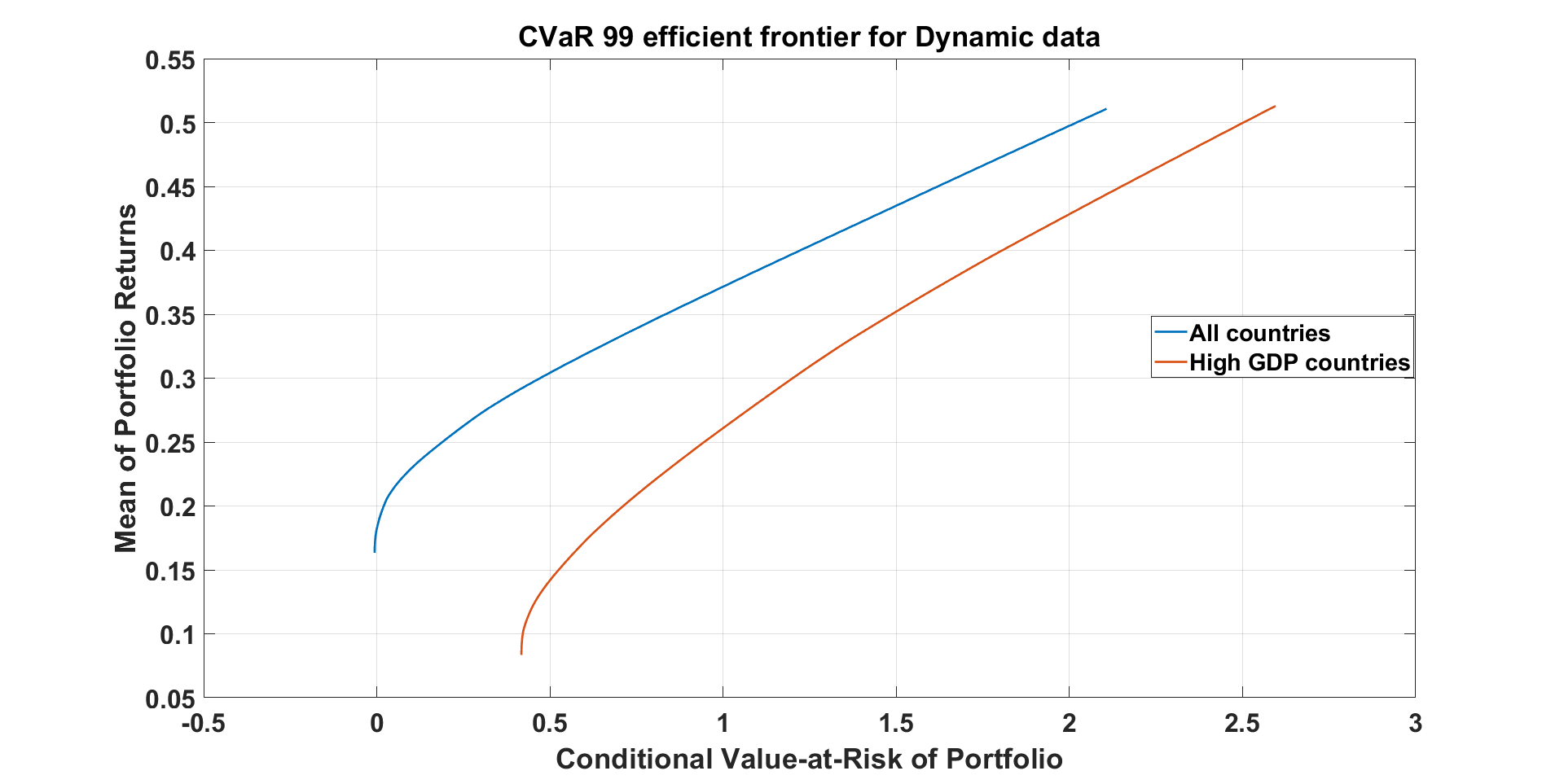}}
	\caption{Comparing Conditional value-at-risk portfolio optimization for dynamic indices: (a) CVaR$_{p,0.05}$ and (b) CVaR$_{p,0.01}$ efficient frontiers.}	
	\label{Fig_EF_CVaR-four}			
\end{figure}

Figures \ref{Fig_EF_CVaR-four}.a and \ref{Fig_EF_CVaR-four}.b reproduce Figure \ref{Fig-EF-Mark-four} for the tail risk measures CVaR$_{p,0.05}$ and CVaR$_{p,0.01}$. The behaviors of the CVaR's EFs of DWI for countries with high GDP compared to those for all countries are comparable. Again, it can be observed that the risk of investing in DWI for all countries is less than for countries with high GPDs at the same expected return. The variation in the behavior of the EFs is more pronounced under the CVaR$_{p,0.01}$ risk measures.

\section{Pricing the options on wellbeing indices} \label{sec:OP}

In this section, we develop a financial model for pricing the DWI options. 
Historical (static) methods cannot deliver proper pricing models on wellbeing indices as they are not in general arbitrage free, which is imperative in option pricing theory.
Thus, we must rely on the no-arbitrage asset pricing theory, and build dynamic models when pricing the wellbeing indices of individual countries and the world.
We use dynamic forecasting models for evaluating future trends of wellbeing indices country by country and globally.
Since the financial options are designed for hedging, risk assessment, and speculating,
we provide DWI option prices for financial institutions to add that additional socioeconomic dimension to their risk-return-adjusted portfolios.

Examples of traditional techniques for calculating the price of options include the Black-Scholes-Merton model, the binomial option pricing model, the trinomial tree, the Monte Carlo simulation, and the finite difference model \citep{hull2003options, duffie2010dynamic}.
We do not employ the Black-Scholes-Merton model to price DWI options due to the presence of heteroskedasticity and distributional heavy-tailedness of the DWI's returns. 
Here, to calculate option prices and explain some well-known mispricing events, the discrete stochastic volatility-based model was developed.\footnote{For different option pricing models, see \cite{black2019pricing, madan1989multinomial, carr2004time, bell2006option, klingler2013option, Shirvanijod_2021_1_138, shirvani2020option} and \cite{shirvani2020optionHu}.} 
In particular, \cite{duan1995garch} proposes the application of discrete-time GARCH to price options. 
We employ the discrete-time GARCH model with NIG innovations to calculate the fair prices of the DWI options while accounting for the accuracy of pricing performance using a volatility-based approach.\footnote{See e.g. \cite{duan1995garch}, \cite{barone2008garch},  and \cite{chorro2012option}.}

In the standard GARCH(1,1) model, \cite{Blaesild_1981} defines the $R_t$ for given $F_{t-1}$, as distributed on real world probability space ($\mathbb{P})$, as follows:
\begin{equation}
\label{DWI_distribution}
R_t \sim NIG\left(\lambda,\frac{\alpha}{\sqrt{a_t}}, \;\frac{\beta}{\sqrt{a_t}}, \;\delta\sqrt{a_t}, \;r'_t+m_t+\mu \sqrt{a_t} \right), \;\;\;\; m_t=\lambda_0 \sqrt{a_t}-\frac{1}{2}a_t .
\end{equation}
\cite{Gerber:1994} describes the traditional approach for determining an equivalent martingale measure to get a constant option price. According to his paper (1994), the Esscher transformation is the conventional method for deriving an equivalent martingale measure to obtain a constant option price.
Since the moment generating function of the NIG distribution has an exponential form, the probability density of $R_t$ is transformed into the risk-neutral probability density using the Esscher transform.

\cite{chorro2012option}  discovered that $R_t$ for given $F_{t-1}$ is distributed  on the risk-neutral probability ($\mathbb{Q}$) using the Esscher transformation as follows:

\begin{equation}
\label{DWI_Q_distribution}
R_t \sim NIG\left(\lambda,\frac{\alpha}{\sqrt{a_t}},\frac{\beta}{\sqrt{a_t}}+\theta_t,\delta \sqrt{a_t},r'_t+m_t+\mu \sqrt{a_t} \right),
\end{equation}
where $\theta_t$ is the solution to $MGF\left(1+\theta_t \right)= MGF\left(\theta_t \right) \, e^{r'_t}$, and $MGF$ is the conditional moment generating function of $R_{t+1}$ given $F_{t}$. \\

\noindent  Using Monte Carlo simulations \citep{chorro2012option}, we construct future values of the DWI to price its call and put options as follows:

\begin{enumerate}
	\item Fitting GARCH(1,1) with NIG innovations to $R_t$ and forecasting $a_1^2$ by setting $t=1$.   
	\item Repeating steps (a)--(d) for $t=3,4,...,T$, where $T$ is time to maturity of the DWI call option from $t=2$.
	\begin{enumerate}
		\item Estimating the model parameter $\theta_t$ using $MGF\left(1+\theta_t \right)= MGF\left(\theta_t \right) \, e^{r'_t}$, where $MGF$ is the conditional moment generating function of $R_{t+1}$ given $F_{t}$ on $\mathbb{P}$. 
		\item Finding an equivalent distribution function for $\epsilon_t$ on $\mathbb{Q}$ and 
		\item Generating the value of $\epsilon_{t+1}$ under the assumption $\epsilon_{t} \sim NIG(\lambda,\alpha,\beta+\sqrt{a_t}\theta_t,\delta,\mu)$ on $\mathbb{Q}$.  
		\item Computing the values of $R_{t+1}$ and $a_{t+1}$ using a GARCH(1,1) model with NIG innovations.
		
	\end{enumerate}

	\item Generating future values of $R_t$ for $t=1,....,T$ on $\mathbb{Q}$ where $T$ is the time to maturity. Recursively, future values of the DWI  are obtained by 
	\begin{equation}
	DWI_{t}= \exp(R_{t}) \cdot DWI_{t-1}.
	\end{equation}
	\item Repeating steps 2 and 3 for 10,000 ($N$) times to simulate $N$ paths to compute future values of the DWI. 
\end{enumerate}

Then, for a specific strike price $K$, the approximate future values of the DWI at time $t$ are the average of  the DWIs, and this price is used to determine the price of the call ($\hat{C}$ and $\hat{P}$, respectively).

\begin{equation} \label{Eq:Call}
\hat{C}\left(t,T,K \right)=\frac{1}{N}\,e^{-r'_t(T-t)}\sum_{i=1}^{N} \max \left(DWI^{(i)}_T-K,0 \right)\ , 
\end{equation}
\begin{equation} \label{Eq:Put}
\hat{P}\left(t,T,K \right)=\frac{1}{N}\,e^{-r'_t(T-t)}\sum_{i=1}^{N} \max \left(K-DWI^{(i)}_T, 0 \right). 
\end{equation}

Call option pricing ($\hat{C}$) help investors in planning to purchase our DWI stocks at a predetermined strike price within a predetermined time frame (time to maturity).

\begin{figure}[h!] 	
	\centering
	\subfigure[]{\includegraphics[width=0.45\textwidth]{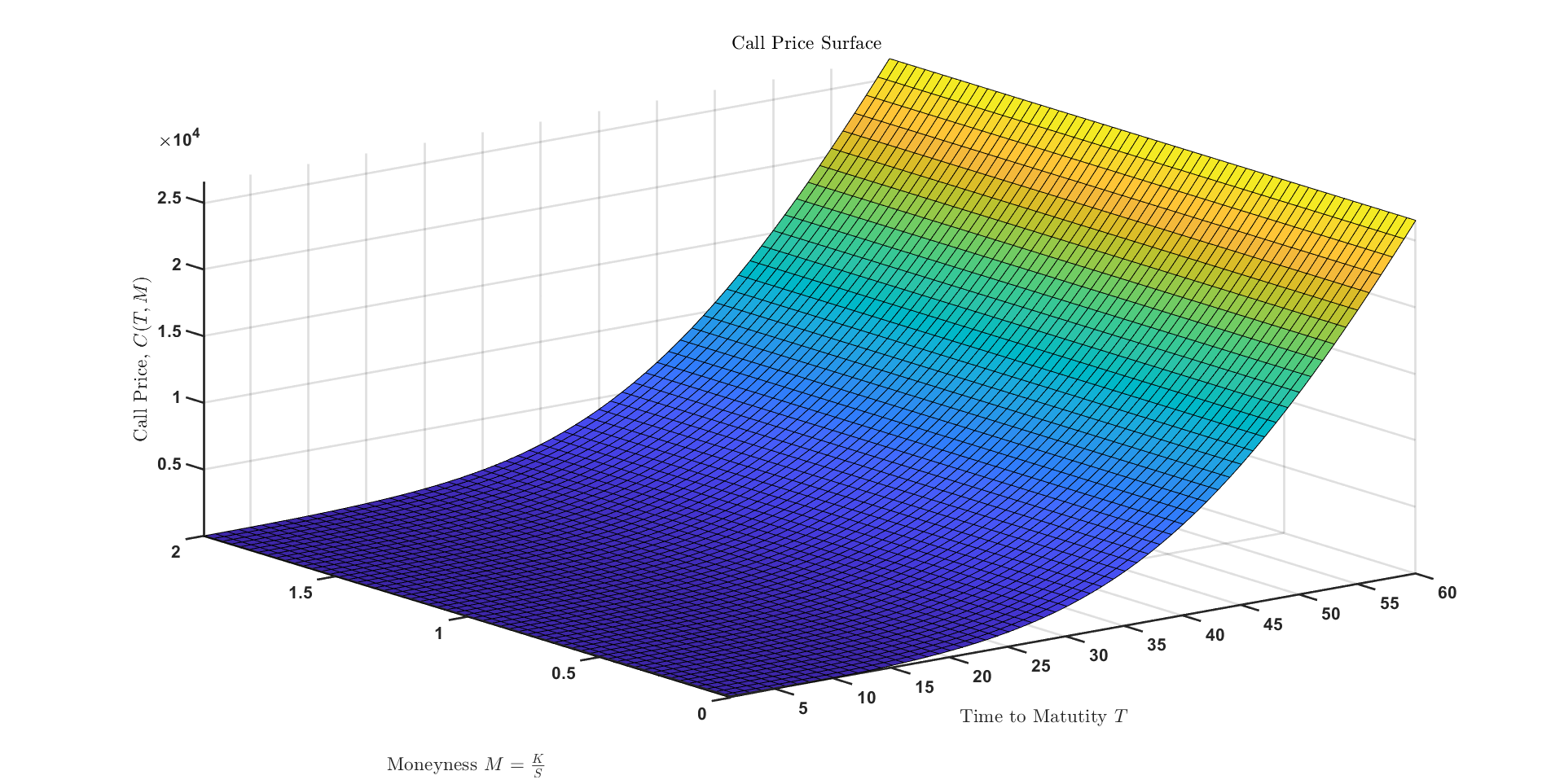}}
	\subfigure[]{\includegraphics[width=0.45\textwidth]{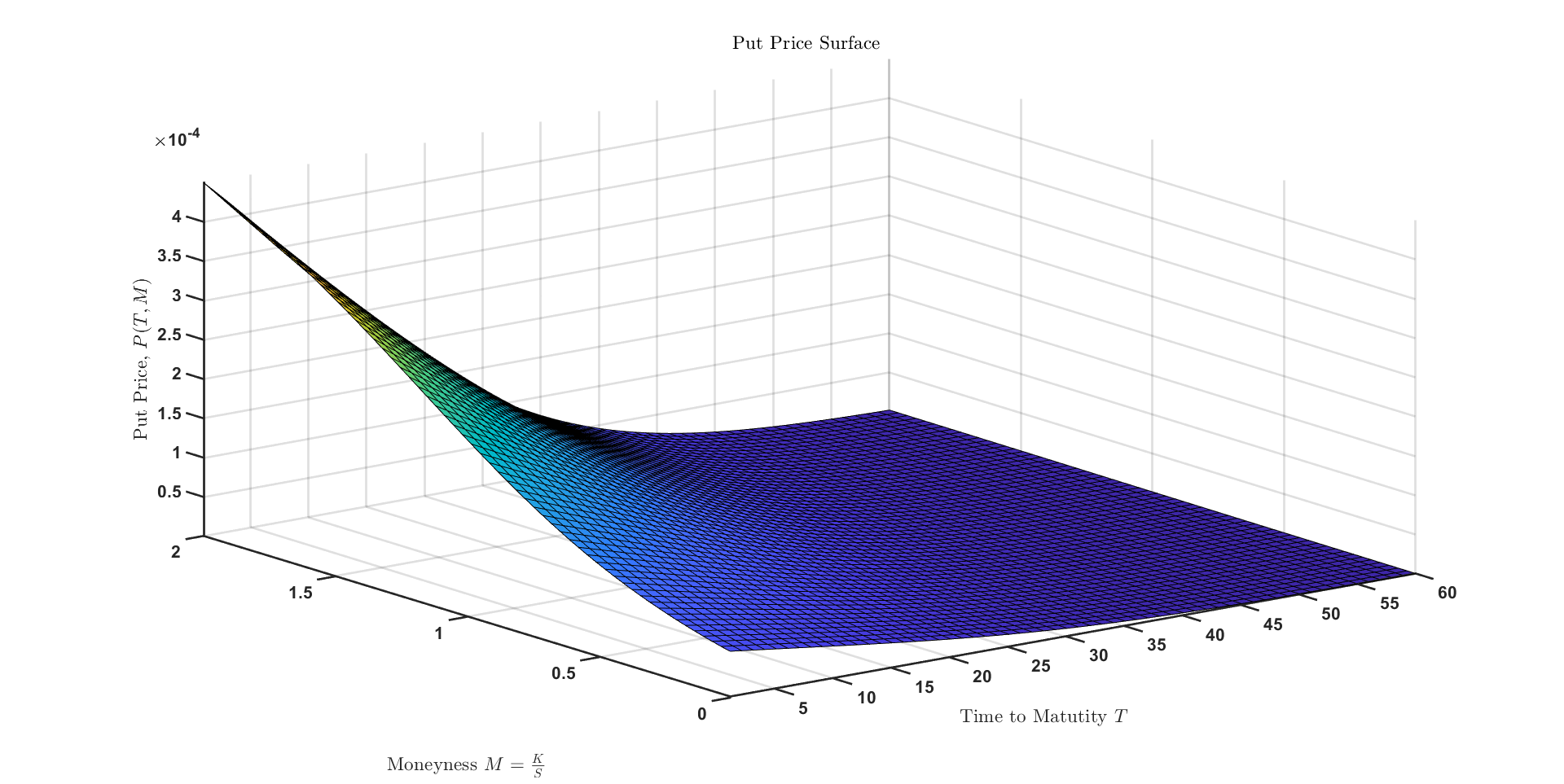}}
	\caption{The option prices for the USA DWI at time $t$ for a given strike price $K$ using a $GARCH(1,1)$ model with NIG innovations. 
 (a) call prices and (b) put prices.}	
	\label{Fig_option_prices}			
\end{figure}

The call and option prices for the DWI are shown in Figure \ref{Fig_option_prices}(a) and depend on the DWI's moneyness $(S/K)$ and time to maturity $(T)$. 
As the strike price rises, the price of the DWI call option slightly decreases. Also, the increase of DWI call option by increasing maturity time represents the increasing price of DWI and consequently the increasing of the price of the population's happiness within the framework of a financial market.

\begin{figure}[h!]
	\centering
	\includegraphics[width=0.75\textwidth]{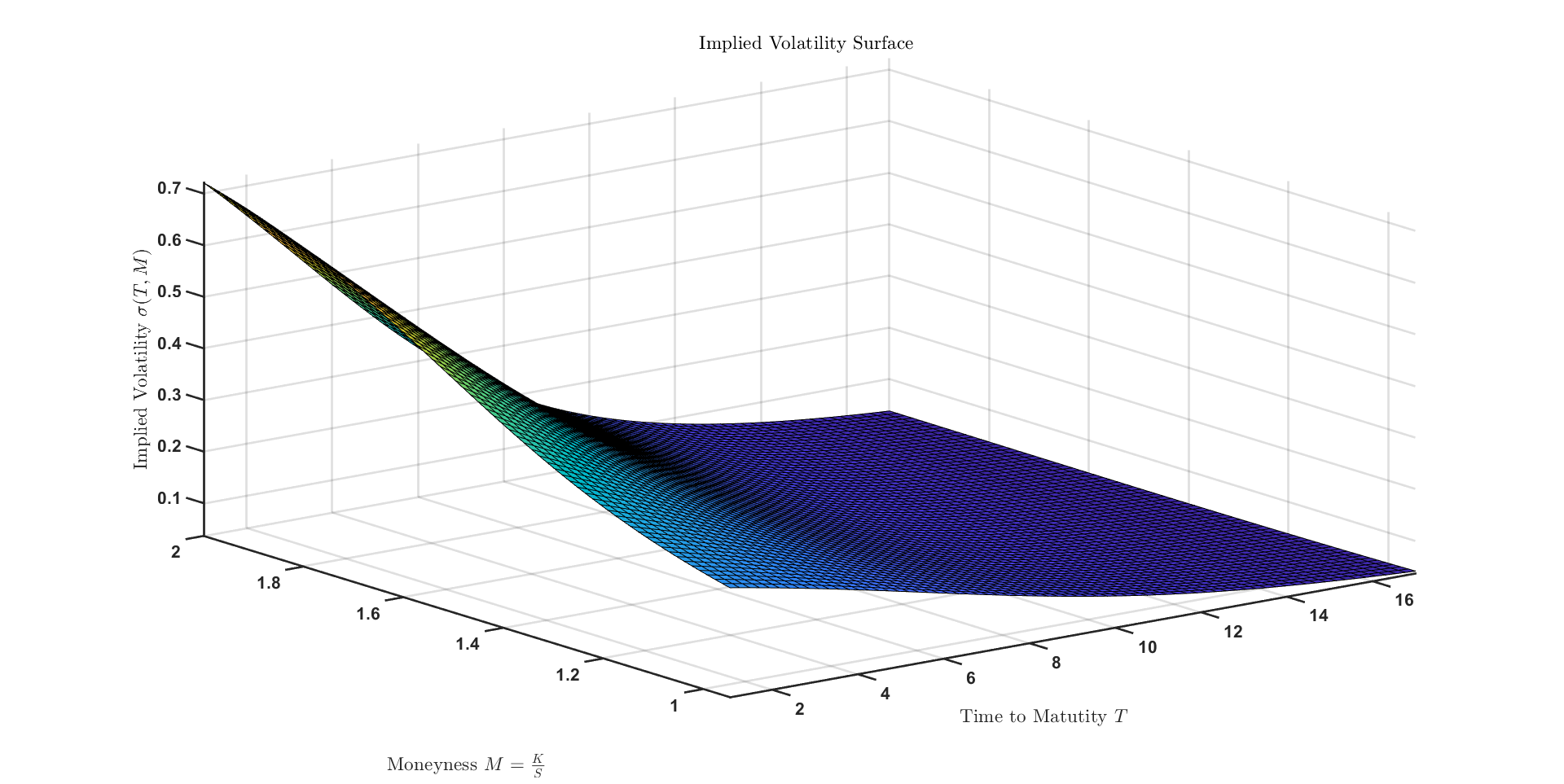} 
	\caption{The USA DWI implied volatilities against time to maturity $(T)$ and moneyness
$(M = \frac{S}{K})$, where $S$ and $K$ are the stock and strike prices, respectively) using a $GARCH(1,1)$ model with NIG innovations}
	\label{Fig_Implied_vol}			
\end{figure}

We exhibit selling prices for the shares in our index in Figure \ref{Fig_option_prices}(b) by explaining the relationship between put option prices ($\hat{P}$) to strike price $(K)$, and time to maturity $(T)$. Figure \ref{Fig_option_prices}(b) reveals that the put option prices are lower than the call option with the same moneyness and time to maturities. However, as expected, the put option price increases as the strike price increases, revealing that the put option price and the strike price may be related linearly. 
Implied volatility is viewed as a reliable forecast of future volatility for the remainder of the option's life.\footnote{See \cite{day1988behavior},  \cite{poterba1984persistence},  \cite{sheikh1989stock}, and \cite{harvey1992market}.}.
Figure \ref{Fig_Implied_vol} displays the implied volatility for DWI using the market values of the call option contracts as a stand-in for the expectation of an upcoming event. The time to maturity ($T$) and moneyness ($M=S/K$, where $S$ and $K$ are the stock and strike prices, respectively) are used to build the implied volatility surface. Typically exhibited during intense market stress, an inverted volatility grin may be seen on the observed volatility surface.
In particular, the highest implied volatilities of the options are observed when the moneyness increases. 

Option implied volatilities are higher for options with lower strike prices than those with higher strike prices. This is because the implied volatility for upside (low strike) equities options is often lower than the implied volatility for equity options in the money, according to the downward sloping (volatility skew) graph. As the period to maturity approaches 16 years, the suggested volatilities tend to converge to a constant.
These option prices should be used for hedging rather than speculation. This instrument should play a role similar to portfolio insurance \citep{portins}.

\nc

\section{Discussion and Conclusion} \label{sec:DC_CI}

We developed a financial market for indices of socioeconomic wellbeing for 
the United States, Australia, Brazil, China, Germany, India, Japan, South Africa, and the United Kingdom
incorporating eight world development socioeconomic indicators.
This has been achieved by creating a global financial market for the wellbeing indices, allowing the financial industry to be involved in assessing, managing, and mitigating (hedging) potential future adverse movements of those indices and the world as a whole.
We introduced and developed the fundamentals of a new financial market model of indices of socioeconomic wellbeing, providing an opportunity for the financial industry to be actively involved in the wellbeing of the societies. 
This paper evaluates the risk of downturns of socioeconomic indices around the world using econometric modeling consistent with dynamic asset pricing theory and creates financial instruments to mitigate those risks.
This allows the financial industry to monitor, manage, and ultimately trade the indices, in this way creating funds for insurance against adverse movements of the indices.

While the ESG financial markets are important components in the financial markets dealing with socioeconomic issues, our goal is more global, encompassing the wellbeing of the societies.
We provided option prices on these indices
which can be used as insurance instruments against adverse movements of wellbeing indices.
Based on asset pricing theory, we derived put option contracts on the indices 
for the financial industry to use in assessing and managing those adverse risks.
Our findings address the question of the total amount of funds needed to improve the global DWI in the future by trading financial contracts representing a variety of insurances against future downturns of the global DWI.

We need more data to prepare an early warning system for downturns of the socioeconomic indices for the countries around the world.
Our wellbeing indices are non-tradable. 
To introduce exchanged trading funds (ETF) \citep{ETF}, we need to replicate the dynamics of the wellbeing indices in terms of fixed income portfolios by applying the tools of asset liability management (ALM) \citep{zenios2007handbook}.
For example, once the US DWI is replicated by an ETF of fixed-income securities, the put options can be viewed as US DWI's insurance instruments.
If US DWI in 2024 is down, then the investors will be paid the put-value on the ETF that has replicated the US DWI.


Our proposed method for constructing indices of socioeconomic  wellbeing is not limited to the nine countries discussed in this paper.
A similar portfolio analysis will be done for various regions in the world based on geographical and economic criteria.
This allows addressing the following problems: which individual or group of countries contribute to the world's wellbeing? and to what degree? 
We plan to extend these financial management principles to construct a financial market of socioeconomic wellbeing for the organization of the petroleum exporting countries (OPEC), by developing the OPEC DWI.
Then, we check the contribution of each OPEC country to the OPEC DWI
and determine what would be the effect if a country leaves or joins OPEC.
We believe these findings will help the financial industry which works with world organizations to address the potential socioeconomic issues related to the wellbeing of the societies.\\

\newpage
\section*{Appendix} \label{sec:Appendix}

\subsection{Robust regression:}

\begin{figure}[h!]
	\centering
	\includegraphics[width=0.75\textwidth]{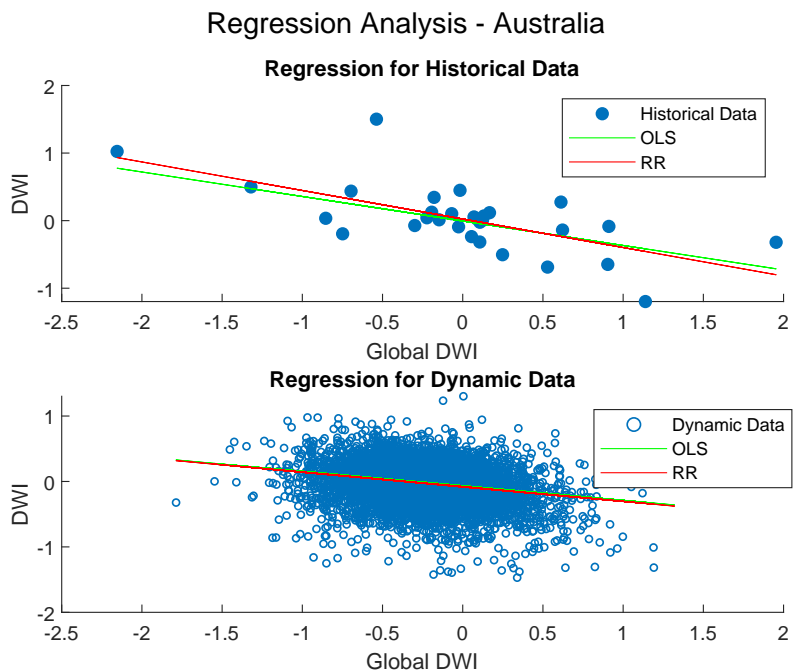} 
	\caption{Robust regression for historical and dynamic log returns in Australia. Regression lines for both historical and dynamic data result in downward forecasts.} 
	\label{Fig-DD_Aus}		
\end{figure}

\begin{figure}[h!]
	\centering
	\includegraphics[width=0.75\textwidth]{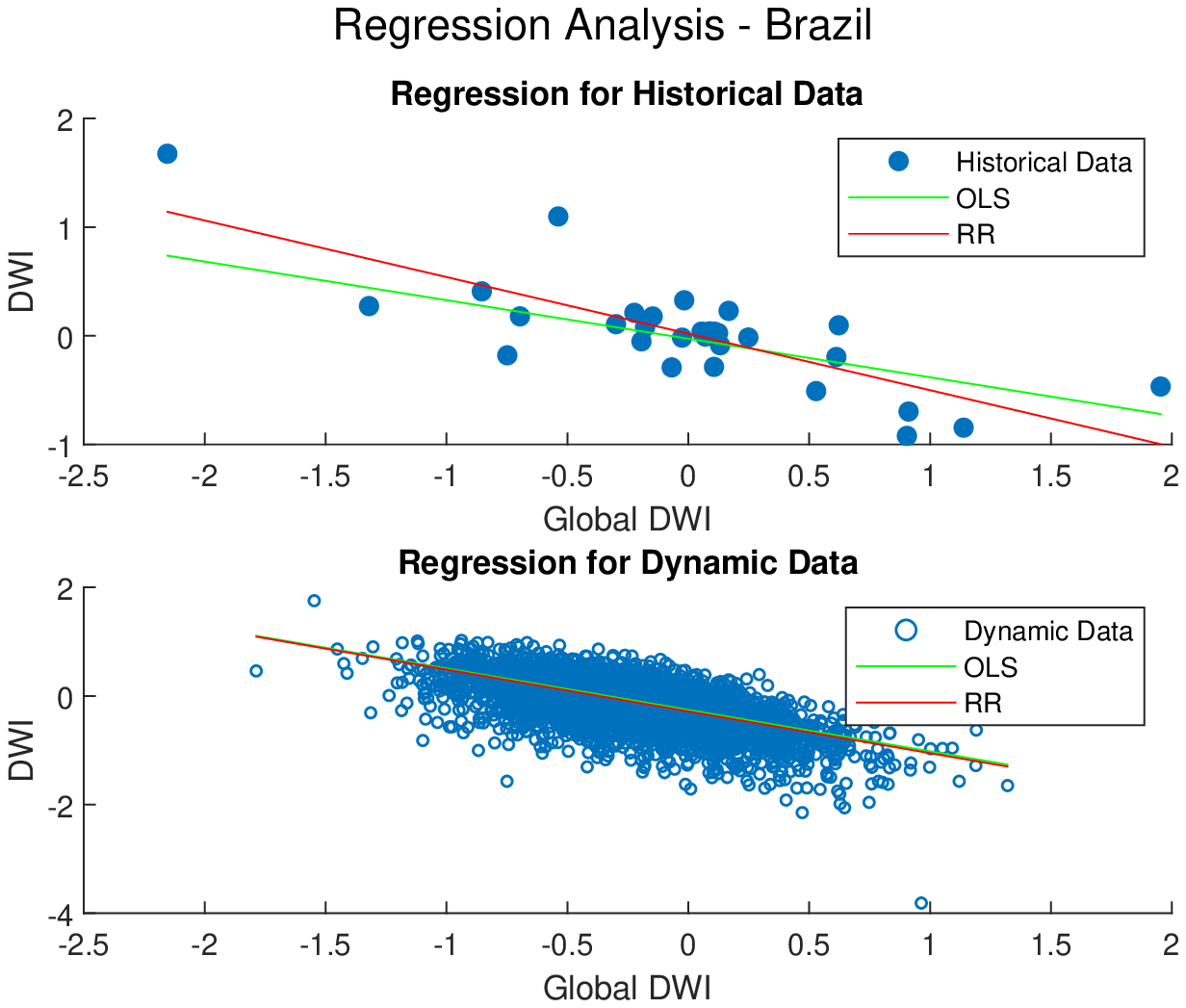} 
	\caption{Robust regression for historical and dynamic log returns in Brazil. Regression lines for both historical and dynamic data result in downward forecasts.} 
	\label{Fig-DD_Bra}			
\end{figure}

\begin{figure}[h!]
	\centering
	\includegraphics[width=0.75\textwidth]{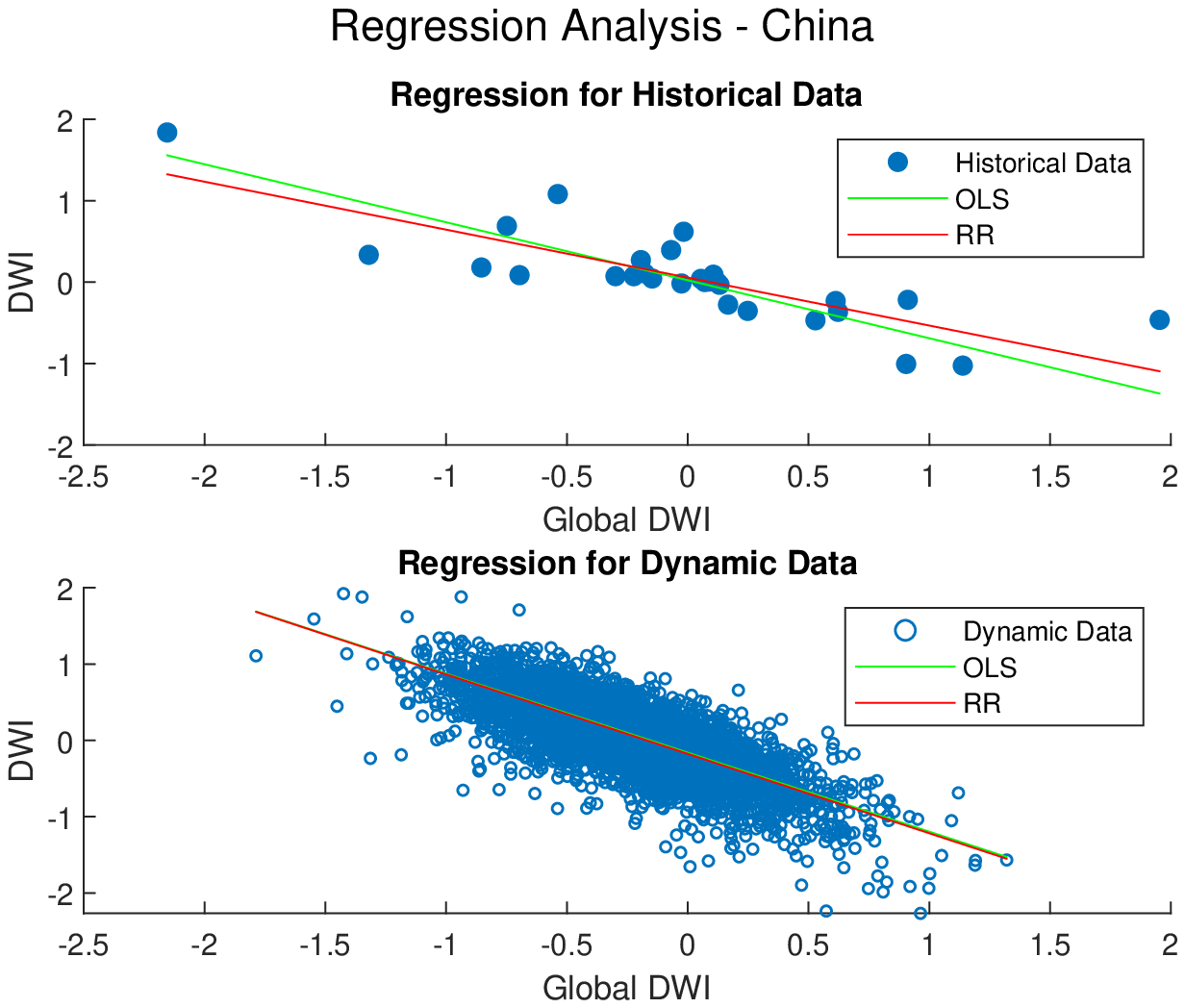} 
	\caption{Robust regression for historical and dynamic log returns in China. Regression lines for both historical and dynamic data result in downward forecasts.} 
	\label{Fig-DD_Chi}			
\end{figure}

\begin{figure}[h!]
	\centering
	\includegraphics[width=0.75\textwidth]{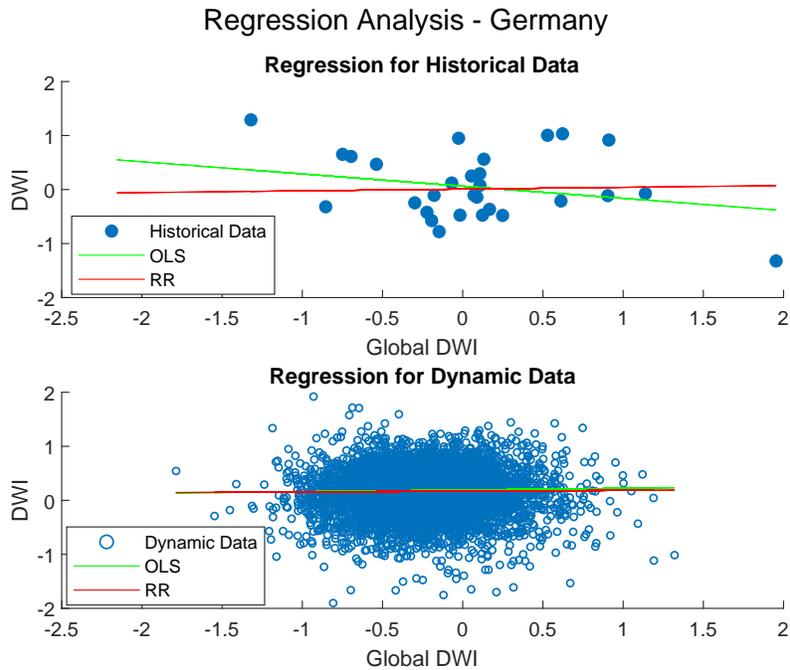} 
	\caption{Robust regression for historical and dynamic log returns in Germany. Regression lines for both historical and dynamic data result in non-significant upward forecasts.} 
	\label{Fig-DD_Ger}			
\end{figure}

\begin{figure}[h!]
	\centering
	\includegraphics[width=0.75\textwidth]{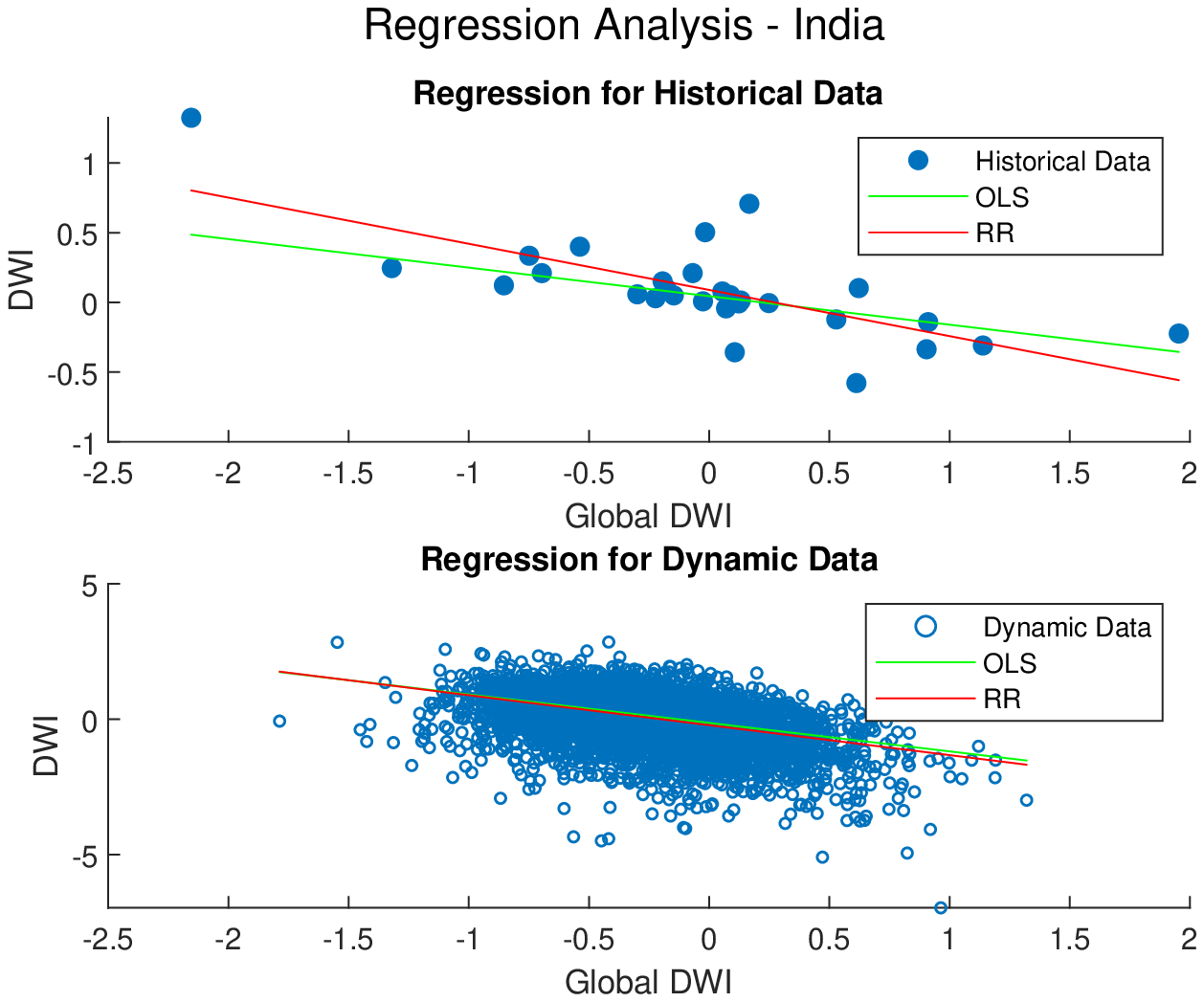} 
	\caption{Robust regression for historical and dynamic log returns in India. Regression lines for both historical and dynamic data result in downward forecasts.} 
	\label{Fig-DD_Ind}			
\end{figure}

\begin{figure}[h!]
	\centering
	\includegraphics[width=0.75\textwidth]{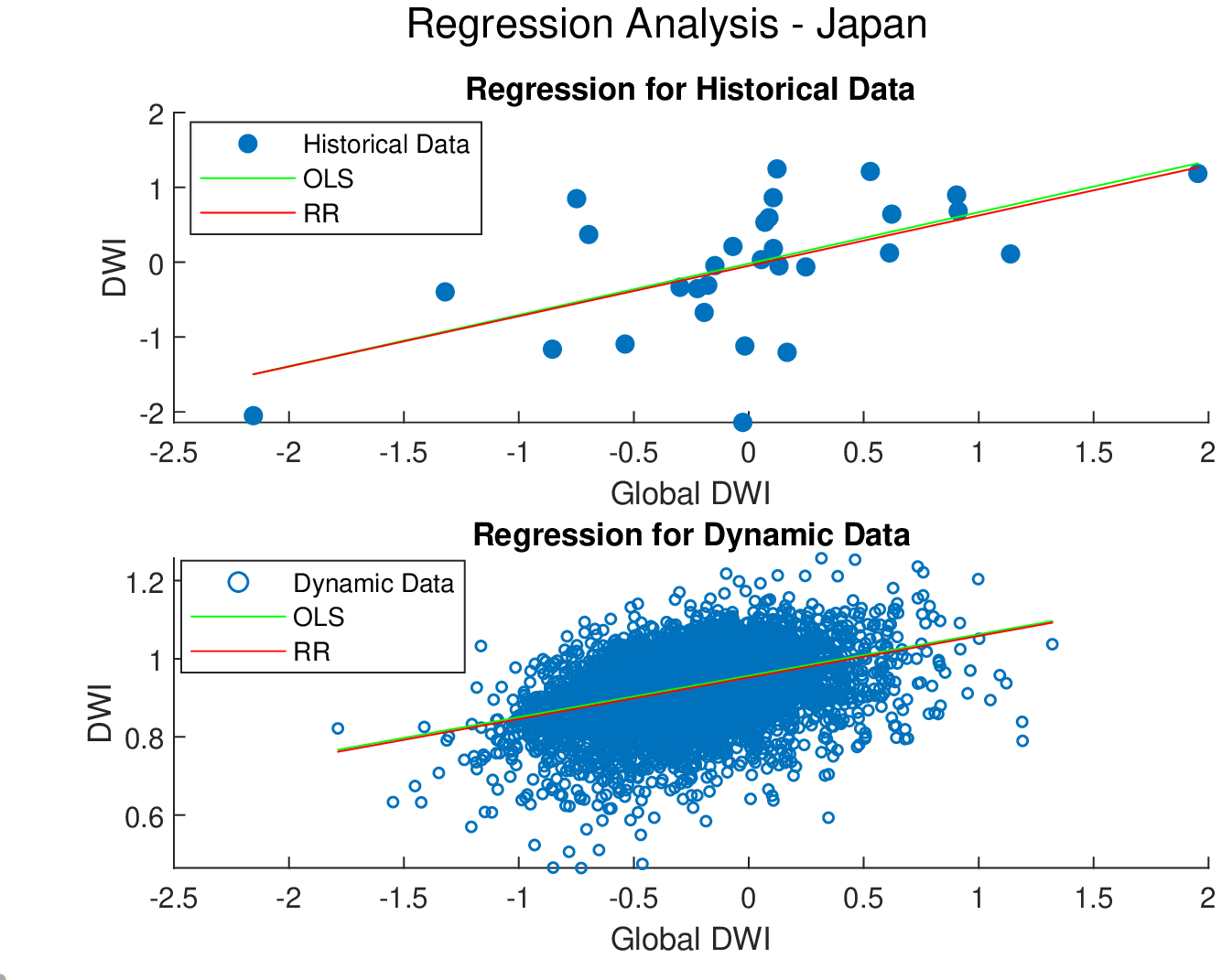} 
	\caption{Robust regression for historical and dynamic log returns in Japan. Regression lines for both historical and dynamic data result in upward forecasts.} 
	\label{Fig-DD_Jap}			
\end{figure}

\begin{figure}[h!]
	\centering
	\includegraphics[width=0.75\textwidth]{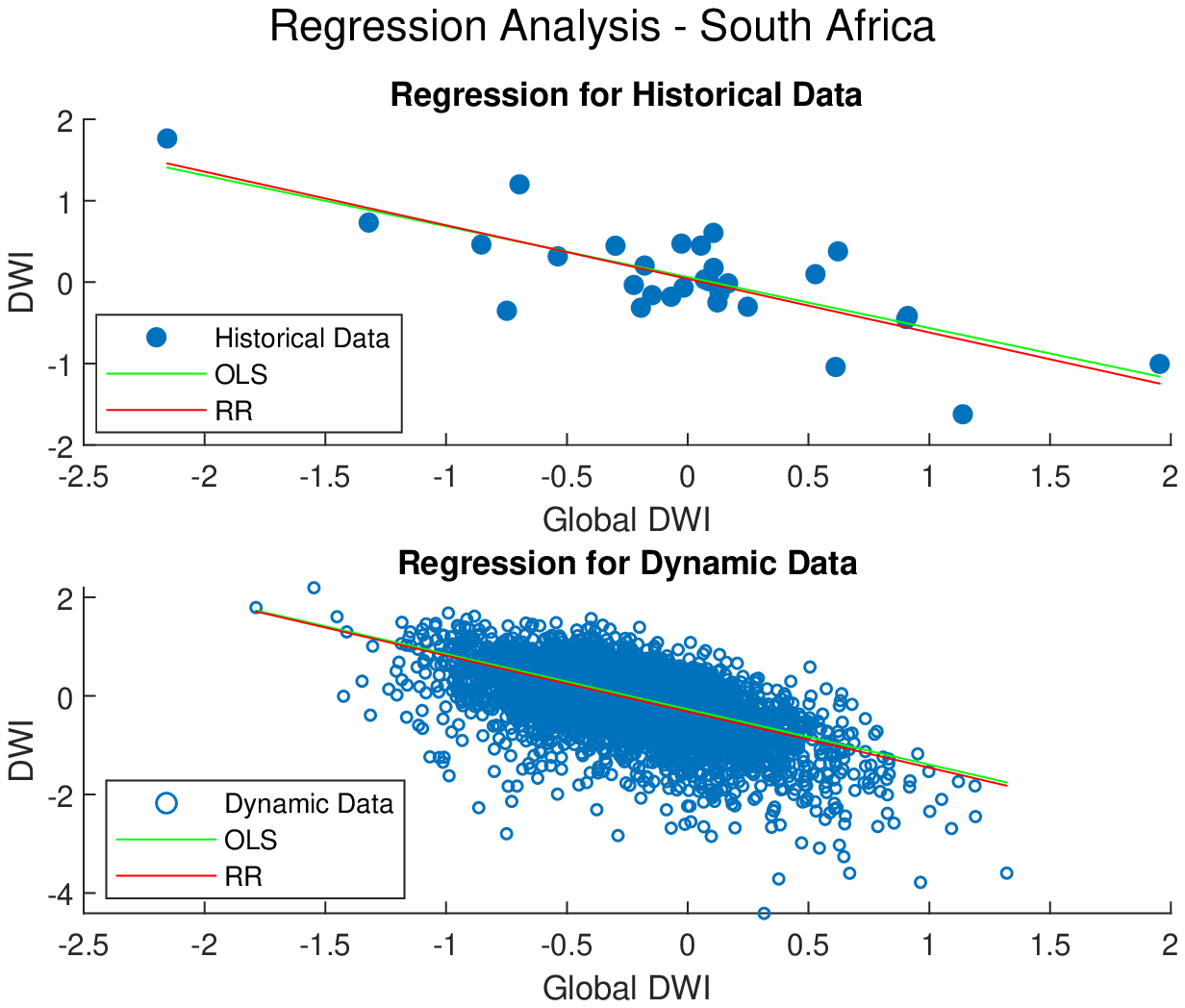} 
	\caption{Robust regression for historical and dynamic log returns in South Africa. Regression lines for both historical and dynamic data result in downward forecasts.} 
	\label{Fig-DD_SA}			
\end{figure}

\begin{figure}[h!]
	\centering
	\includegraphics[width=0.7\textwidth]{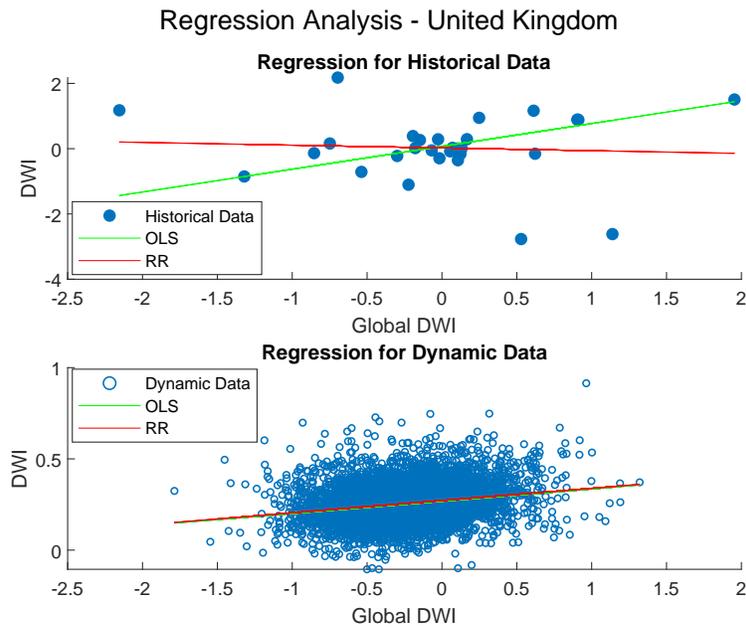} 
	\caption{Robust regression for historical and dynamic log returns in the United Kingdom. Regression lines historical data does not yield a significant downward trend and dynamic data  does not yield a significant upward trend.} 
	\label{Fig-DD_UK}			
\end{figure}
\clearpage
\normalem

\end{document}